\documentclass[10pt,a4paper]{article}
\usepackage[hmargin=1in, vmargin=1in]{geometry}

\usepackage{titling}

\usepackage[utf8]{inputenc} 
\usepackage[T1]{fontenc}    
\usepackage{hyperref}     
\usepackage{url}            
\usepackage{booktabs}       
\usepackage{amsfonts}       
\usepackage{nicefrac}       
\usepackage{microtype}      
\usepackage{graphicx}
\usepackage{doi}

\usepackage{amsmath}
\usepackage{amsfonts}
\usepackage{amssymb}
\usepackage{bm}
\usepackage{booktabs,multirow,multicol,makecell}
\usepackage[font=small,labelfont=small]{caption}
\usepackage{dsfont}
\usepackage{enumitem}
\usepackage{mathtools}
\usepackage{nicefrac}
\usepackage{parskip}
\usepackage{placeins}
\usepackage{newunicodechar,textgreek}
\usepackage{svg}
\usepackage{titletoc}
\usepackage{xcolor}

\usepackage{algorithm}
\usepackage{algpseudocode}

\usepackage[backend=biber,
            eprint=false,
            giveninits=true,
            maxcitenames=2,
            mincitenames=1,
            maxbibnames=10,
            minbibnames=1,
            sorting=none,
            style=phys,
            uniquename=init,
            url=false]{biblatex} 
\DeclareNameAlias{author}{family-given}
\AtEveryBibitem{\clearlist{language}}
\AtEveryBibitem{\clearfield{month}}
\AtEveryBibitem{\clearfield{note}}

\usepackage{tabu}
\newcommand{\tblspace}{\vspace{0em}}

\usepackage{authblk}

\usepackage{mycommands}

\graphicspath{ {./figures/} }

\definecolor{linkcolour}{rgb}{0.0, 0.2, 0.6}
\hypersetup{
  colorlinks=true,
  breaklinks,
  urlcolor=linkcolour,
  linkcolor=linkcolour,
  citecolor=linkcolour
}

\renewcommand{\eqref}[1]{Eq.~\textup{(\ref{#1})}}

\title{Reliable model selection in the presence of parameter non-identifiability}
\author[1,2]{Yong See Foo}
\author[3]{Torkel E. Loman}
\author[1]{Alexander P. Browning}
\author[4]{Ivo Siekmann}
\author[3]{Ruth E. Baker}
\author[1,2]{Jennifer A. Flegg}
\affil[1]{School of Mathematics and Statistics, University of Melbourne, Australia}
\affil[2]{ARC Centre of Excellence for the Mathematical Analysis of Cellular Systems, University of Melbourne, Australia}
\affil[3]{Wolfson Centre for Mathematical Biology, Mathematical Institute, University of Oxford, United Kingdom}
\affil[4]{School of Computer Science and Mathematics, Liverpool John Moores University, United Kingdom}

\date{}

\begin{document}

\maketitle

\begin{refsection}

\begin{abstract}
  Mathematical models are invaluable for understanding and predicting how biological systems behave, although their construction requires specifying mechanisms and relationships that are often not perfectly known. In the presence of multiple competing models, model uncertainty should be accounted for when performing inference based on available data. Bayesian model selection is a framework for testing mechanistic hypotheses and generating predictions under model uncertainty, which generally requires computation of the model evidence. In this work, we investigate the reliability of evidence computation methods when parameter non-identifiability---the inability to distinguish between parameter values given available data---is present, and find that deterministic evidence approximations can produce misleading model selection results because their underlying assumptions are violated. We propose a novel implementation of adaptive multiple importance sampling for evidence estimation, and demonstrate its robustness against non-identifiability. We use ecological case studies to demonstrate how simple model selection methods fail to produce accurate results, whereas our method yields model selection results that are comparable to those obtained by Markov chain Monte Carlo methods at substantially lower computational cost. Given the pervasiveness of parameter non-identifiability in mathematical biology, this work provides a practical approach to reliable model selection in the presence of poorly identified parameters.
\end{abstract}

\renewcommand{\topfraction}{.99}
\renewcommand{\textfraction}{.01}
\renewcommand{\floatpagefraction}{.99}

\section{Introduction}

Mechanistic models in mathematical biology are simplified abstractions of complex biological systems, where the processes involved are often not fully understood.  Multiple competing models are often proposed to formalize different hypotheses about the underlying mechanisms and relationships that govern biological dynamics. Selecting between such models is therefore a fundamental task in mathematical biology~\cite{burnhamAICModel2011,kirkModelSelection2013}. The most widespread model selection approach is to rank models via information criteria, such as the Akaike information criterion (AIC)~\cite{akaikeNewLook1974} and Bayesian information criterion (BIC)~\cite{schwarzEstimatingDimension1978}, due to their computational simplicity. The AIC and BIC have been used to select models across a wide range of applications---including ecology~\cite{linkModelWeights2006}, cancer biology~\cite{lalehClassicalMathematical2022}, epidemiology~\cite{stocksModelSelection2020}, and cell biology~\cite{warneUsingExperimental2019}---and there is substantial theoretical discussion about the relative merits of different information criteria~\cite{ahoModelSelection2014,dingModelSelection2018,zhangInformationCriteria2023}.

Model selection is not limited to selecting a single model. In fact, if multiple models can plausibly explain some observed data at hand, drawing conclusions that solely rely on a single ``best'' model ignores model uncertainty~\cite{draperAssessmentPropagation1995,kassBayesFactors1995}. For example, if multiple models that each explain the data well produce diverging predictions, one can arrive at overconfident conclusions if the predictions of only one model are considered. Bayesian model selection provides a principled framework to perform model averaging~\cite{leamerSpecificationSearches1978,hoetingBayesianModel1999}, which combines inference results and ensuing predictions obtained over multiple candidate models. Model averaging has gained increasing attention in ecological modeling~\cite{hootenGuideBayesian2015,dormannModelAveraging2018}, though applications are also present in other fields such as biochemistry~\cite{vyshemirskyBayesianRanking2008}, epidemiology~\cite{abboudForecastingPathogen2023}, and systems biology~\cite{linden-santangeliIncreasingCertainty2025}.

Bayesian model selection generally requires calculating the model evidence of each candidate model, which is fundamentally a problem of numerical integration. Specifically, the model evidence is the integral of the unnormalized posterior over parameter space. The simplest approach is to approximate the model evidence using the BIC~\cite{rafteryBayesianModel1995}. The BIC is commonly used to select and rank models without necessarily interpreting it as an approximation of model evidence. Crucially, the derivation of the BIC relies on the assumption that the dataset is sufficiently large, an assumption that is not always met by practical applications in mathematical biology. Alternatively, there are Monte Carlo methods for estimating model evidence that do not rely on large-data assumptions; see~\cite{llorenteMarginalLikelihood2023} for a recent review. Monte Carlo integration methods operate by treating integrals as expectations, which are estimated by taking sample means. Importance sampling~\cite{robertMonteCarlo2004} underpins a broad class of Monte Carlo approaches. It operates by drawing samples from a simple-to-evaluate proposal distribution while accounting for the discrepancy between the posterior distribution and the proposal distribution. The efficiency of importance sampling estimates depends on how well the proposal distribution approximates the posterior distribution. Methods such as reverse importance sampling~\cite{gelfandBayesianModel1994} and bridge sampling~\cite{mengSimulatingRatios1996a} assume the availability of samples from the posterior distribution; these methods are less commonly used as sampling from the posterior distribution of each model can be computationally intensive. There are also posterior sampling methods that estimate the model evidence as a byproduct, e.g., Sequential Monte Carlo~\cite{delmoralSequentialMonte2006} and nested sampling~\cite{skillingNestedSampling2006}, though these methods may require high computational costs to deliver accurate evidence estimates~\cite{snyderObstaclesHighdimensional2008,beskosErrorBounds2014,dittmannNotesPractical2024}.

At the same time, there is a growing body of literature in mathematical biology concerning the issue of parameter non-identifiability~\cite{gutenkunstUniversallySloppy2007,raueStructuralPractical2009,wielandStructuralPractical2021}. Parameter non-identifiability describes the scenario in which available data is insufficiently informative in constraining the parameter values of a model. Literature on parameter identifiability often distinguishes between structural non-identifiability and practical non-identifiability; the former is concerned with non-identifiability arising from the model structure, while the latter is concerned with non-identifiability arising from insufficient information present in the available data~\cite{raueStructuralPractical2009}. Identifiability considerations have been proposed as a criterion for model selection, with some authors advocating the prioritization of models with identifiable parameters~\cite{simpsonParameterIdentifiability2022,liuParameterIdentifiability2024,wangBayesianIdentifiability2026}. In this work, we investigate a different question concerning identifiability and model selection: are common model selection methods reliable when applied in the presence of practical non-identifiability? Given that non-identifiability can cause numerical and convergence issues for statistical inference~\cite{raueJoiningForces2013,mullerMarkovChainMonteCarlo2018}, it is important to understand how non-identifiability affects the reliability of model selection methods. Moreover, we expect non-identifiability to be prevalent for model selection problems that involve over-parameterized models, i.e. models that contain more parameters than can be meaningfully constrained by data.

To investigate the impact of parameter identifiability on Bayesian model selection, we compare BIC and Monte Carlo approaches for calculating model evidence, and evaluate how errors in evidence estimation propagate to model selection results. Simple approaches such as the BIC rely on the assumption that the posterior is approximately Gaussian (in the case of BIC, this is implied by its large-data assumption). Such an assumption is unlikely to hold when parameters are non-identifiable, as non-identifiability is often associated with ridge-shaped posterior distributions. In response to identifiability concerns, we propose a novel evidence computation method based on adaptive multiple importance sampling (AMIS)~\cite{cornuetAdaptiveMultiple2012}. This method aims to improve the robustness of AMIS against parameter non-identifiability, as standard AMIS may face difficulties when adapting a proposal distribution to a posterior distribution that is non-elliptical due to non-identifiable parameters. Using bridge sampling as a gold standard, we evaluate the reliability and efficiency of evidence computation methods when performing model selection for differential equation applications in the presence of parameter non-identifiability.

\section{Statistical preliminaries}

We provide a brief overview of single-model Bayesian inference, and describe Bayesian model selection as a framework for performing multi-model inference.

\subsection{Bayesian inference}

Given some observed data~$\data$, suppose we propose a model~$\model$ to explain the data. In this work, we focus on scenarios where the data consists of time-course observations of a dynamical system, while the model describes the underlying dynamics with parameterized differential equations, the distribution of the model parameters without knowledge of the data, and the noise associated with the observations. Denoting the parameters as $\param$, which we assume to be continuous and unconstrained, the \emph{likelihood} $p(\data\vert\param,\model)$ describes the dynamics and noise distribution under model~$\model$, while the \emph{parameter prior} $p(\param\vert\model)$ describes the distribution of the parameters without knowledge of the data under model~$\model$. The likelihood and parameter prior represent distributions that are model-dependent, though this dependence is not always notationally explicit.

The primary goal of Bayesian inference is to compute the \emph{parameter posterior} $p(\param\vert\data,\model)$:
\begin{equation}\label{eq:param_post}
    p(\param\vert\data,\model) = \frac{p(\data\vert\param,\model) p(\param\vert\model)}{\int p(\data\vert\param,\model) p(\param\vert\model) \, \dd{\param}}.
\end{equation}
 The parameter posterior distribution quantifies inference about parameters $\param$ upon observing data~$\data$ under model~$\model$. In practice, the parameter posterior $p(\param\vert\data,\model)$ is not evaluated directly, as the integral in~\eqref{eq:param_post} is generally not available in analytical form. Instead, one typically obtains samples $\{\param_n\}_{n=1}^N$ of the parameter posterior, for example, using Markov chain Monte Carlo (MCMC) methods~\cite{tierneyMarkovChains1994}. 
 
 Bayesian estimation of some quantity of interest $f(\param,\model)$ (e.g., a forecasting prediction) under model~$\model$ amounts to obtaining the posterior distribution of $f(\param,\model)$, which we denote as $p(f \vert \data, \model)$. Given posterior samples $\{\param_n\}_{n=1}^N$ drawn from the parameter posterior $p(\param\vert\data,\model)$, one typically approximates $p(f \vert \data, \model)$ with the empirical distribution of $\{f(\param_n,\model)\}_{n=1}^N$, e.g., via kernel density estimation.


\subsection{Bayesian model selection}\label{sec:model_selection}


Bayesian model selection is a framework for comparing and combining single-model inference results obtained from multiple candidate models. Suppose we have a set of $\nmodel$ candidate models that are proposed to explain the data~$\data$. We denote this model set as $\modelspace = \{\model_1,\ldots,\model_\nmodel\}$. The key object in Bayesian model selection is the \emph{model posterior} $p(\model \vert \data)$:
\begin{equation}\label{eq:model_post}
    p(\model \vert \data) = \frac{p(\data\vert\model)p(\model)}{\sum_{m=1}^M p(\data\vert\model_m)p(\model_m)},
\end{equation}
where $p(\model)$ is the prior probability of model~$\model$ and $p(\data\vert\model)$ is the \emph{model evidence} (also known as the marginal likelihood) of $\model$, defined as
\begin{equation}\label{eq:evidence}
    p(\data\vert\model) = \int p(\data\vert\param,\model) p(\param\vert\model) \, \dd{\param},
\end{equation}
which coincides with the denominator in \eqref{eq:param_post}. The model posterior in \eqref{eq:model_post} can be understood as a model analogue of the parameter posterior in \eqref{eq:param_post}. The model evidence plays the role of a likelihood in \eqref{eq:model_post} to perform inference over models.

A common use of model posterior probabilities is pairwise model comparison via Bayes factors~\cite{kassBayesFactors1995}, that is, to compare models $\model,\model'\in\modelspace$ using the odds ratio ${p(\model \vert \data)/p(\model' \vert \data)}$. Another use of the model posterior distribution is in \emph{Bayesian model averaging} (BMA)~\cite{hoetingBayesianModel1999}, which combines single-model Bayesian inference results by using model posterior probabilities as weights. To illustrate BMA, suppose we seek to obtain the posterior distribution of some quantity of interest $f(\param, \model)$ while considering all models in $\modelspace$. Recall that in the single-model case, we can find the posterior distribution of $f$ under some model~$\model$, namely $p(f\vert\data,\model)$. In the multi-model case, the model-averaged posterior distribution of $f$ is the model posterior-weighted mixture of single-model posterior distributions, namely
\begin{equation}
    p(f \vert \data, \modelspace) = \sum_{\model \in \modelspace} p(\model \vert \data) p(f \vert \data, \model).
\end{equation}
As a simple example, suppose a subset $\modelspace^* \subset \modelspace$ of models include some biological feature of interest. Let $f = 1$ if model~$\model$ includes this feature, otherwise $f = 0$. The model-averaged posterior probability that this biological feature is present is given by $\sum_{\model \in \modelspace^*} p(\model \vert \data)$. As another example, the model-averaged posterior mean of a forecasting prediction $f^\text{pred}$ is the model posterior-weighted average of single-model posterior means $\bar{f}^\text{pred}_\model$ of the forecasting prediction under each model~$\model\in\modelspace$, namely $\sum_{\model\in\modelspace} p(\model \vert \data) \bar{f}^\text{pred}_\model$.

\section{Methods for computing model evidence}

A key challenge in Bayesian model selection is the computation of the model evidence as used in~\eqref{eq:model_post}, which generally are analytically intractable integrals. In this section, we discuss methods for computing the model evidence $p(\data\vert\model)$ for a model~$\model$, letting $Z$ denote the model evidence $p(\data\vert\model)$, $\trg(\param)$ denote the parameter posterior distribution $p(\param \vert \data,\model)$, and $\unnormtrg(\param)$ denote the unnormalized posterior $p(\data\vert\param,\model)p(\param\vert\model)$. With this notation, we have that $\trg = \unnormtrg/Z$ and rewrite~\eqref{eq:evidence} as
\begin{equation}\label{eq:evidence_rewrite}
Z = \int \unnormtrg(\param) \, \dd{\param}.
\end{equation}
We assume that the unnormalized posterior $\unnormtrg$ can be numerically evaluated, though a large number of evaluations can be computationally expensive, e.g., when likelihood evaluation requires numerically solving differential equations.

\subsection{Deterministic approximations of evidence}\label{sec:bic}

Model selection is commonly performed by evaluating information criteria. Here, we discuss the BIC and the Laplace approximation, which are deterministic approximations of the model evidence. The Laplace approximation is derived by applying a second-order Taylor approximation to the log integrand in~\eqref{eq:evidence_rewrite}. Let $\MAP$ denote the maximizer of $\log \unnormtrg(\param)$, known as the \textit{maximum a posteriori} estimate, and let $\MAPhess$ denote the Hessian matrix of $-\log \unnormtrg(\param)$ at $\MAP$. Assuming that $\MAPhess$ is positive definite, the Laplace approximation approximates the model evidence as
\begin{align}
    Z 
    &= \int \exp \{ \log \unnormtrg(\param) \} \, \dd{\param} \nonumber\\
    &\approx \int \exp \!\left\{ \log \unnormtrg(\MAP) - \frac{1}{2}(\param-\MAP)^\top\MAPhess(\param-\MAP) \right\} \, \dd{\param} \nonumber\\
    &= \unnormtrg(\MAP) \sqrt{(2\pi)^d \text{det}(\MAPhess^{-1})} , \label{eq:Z_laplace}
\end{align}
where $d$ is the dimensionality of $\param$. Under regularity assumptions, the relative error incurred by the approximation in~\eqref{eq:Z_laplace} is of order $\bigO{\ndata^{-1}}$ as $\ndata\rightarrow\infty$, where $\ndata$ is the number of data observations~\cite{tierneyAccurateApproximations1986a}. Taking logarithms of~\eqref{eq:Z_laplace} gives
\begin{align}
    \log Z 
    &= \log \unnormtrg(\MAP) + \frac{d}{2} \log 2\pi + \frac{1}{2} \log \text{det}(\MAPhess^{-1}) + \bigO{\ndata^{-1}} \nonumber\\ 
    &= \log p(\data\vert\MAP,\model) + \log p(\MAP\vert\model) + \frac{d}{2} \log 2\pi - \frac{1}{2} \log \text{det}(\MAPhess) + \bigO{\ndata^{-1}}. \label{eq:logZ_laplace}
\end{align}
To derive the standard BIC expression, we further assume that $\ndata$ is large, the MAP estimate is similar to the maximum likelihood estimate $\MLE$, and $\MAPhess$ scales as $\bigO{\ndata}$ as $\ndata\rightarrow\infty$. Under these assumptions, quantities that are of order $\bigO{1}$ are ignored since the leading term of $\log \text{det}(\MAPhess)$ is $d \log\, \ndata$, producing the approximation 
\begin{equation}\label{eq:const_error}
    \log Z \approx \log p(\data\vert\MLE,\model) - \frac{d}{2} \log\, \ndata.
\end{equation}
Multiplying the expression in~\eqref{eq:const_error} by $-2$ gives the standard BIC expression, $\textrm{BIC}=-2\log p(\data\vert\MLE,\model) + d \log\, \ndata$. See~\cite{draperAssessmentPropagation1995} for further details on the derivation of BIC.

Several works~\cite{linkModelWeights2006,kirkModelSelection2013,hootenGuideBayesian2015} have proposed the use of BIC to approximate model posterior probabilities via the evidence approximation in~\eqref{eq:const_error}. However, the derivation above exposes several ways for this approximation to be inaccurate, especially when parameters are non-identifiable. Firstly, the omission of quantities that are of order $\bigO{1}$ relies on the assumption that $\log\,\ndata$ is large, which is often not the case for biological applications where experimental observations are limited~\cite{lillacciParameterEstimation2010,wielandStructuralPractical2021}. Moreover, since practical non-identifiability is typically associated with near-zero eigenvalues of the Hessian $\MAPhess$~\cite{rothenbergIdentificationParametric1971,gutenkunstUniversallySloppy2007}, ignoring the log eigenvalues can compromise the accuracy of evidence estimation using BIC. These two concerns can be alleviated by instead using the Laplace approximation in~\eqref{eq:Z_laplace}, though we note that both the BIC and Laplace approximation rely on a second-order Taylor approximation, in other words, these methods rely on the posterior $\trg(\param)$ being approximately Gaussian, centered at $\MAP$. When parameters are non-identifiable, the posterior typically exhibits non-Gaussian features such as funnels and ridges; see Figure~\ref{fig:logistic_richards}B and~\cite{raueJoiningForces2013,hinesDeterminationParameter2014} for examples. These concerns motivate the use of Monte Carlo methods, whose accuracy depends on the number of Monte Carlo samples---which the practitioner has control over---instead of the number of observations.

\subsection{Evidence estimation using importance sampling}\label{sec:IS}

\emph{Importance sampling} underpins a large class of Monte Carlo methods that estimate integrals by drawing samples from a proposal distribution and weighting them to account for the discrepancy between the proposal and the posterior. The proposal should be an approximation of the posterior that is computationally cheap to draw samples from compared to drawing samples from the posterior distribution, e.g., via MCMC. Importance sampling is based on the following identity:
\begin{equation*}
    Z 
    = \int \unnormtrg(\param) \, \dd{\param} 
    = \int q(\param) \frac{\unnormtrg(\param)}{q(\param)} \, \dd{\param}  
    = \E[q]{\frac{\unnormtrg(\param)}{q(\param)}},
\end{equation*}
which holds for any proposal distribution $q$ whose support contains that of $\trg$, where $\mathbb{E}_{q}$ denotes expectation with respect to $q$. In practice, given $N$ samples $\{\param_n\}_{n=1}^N$ drawn from a proposal~$q$, importance sampling produces an unbiased estimator of the evidence $Z$, namely
\begin{equation}\label{eq:IS_estimator}
    \hat{Z} = \frac{1}{N} \sum_{n=1}^N \frac{\unnormtrg(\param_n)}{q(\param_n)}.
\end{equation}
To ensure finite variance of the estimator $\hat{Z}$, we require the tails of $q$ to not be lighter than those of $\trg$~\cite{robertMonteCarlo2004}. In other words, $q$ must place probability mass wherever $\trg$ does for importance sampling to reliably estimate the evidence. Therefore, despite the unbiasedness of the importance sampling estimator, its finite-sample accuracy depends on how well the proposal covers the posterior.

One can use the Laplace approximation to inform a simple choice of the proposal~$q$~\cite{kukLaplaceImportance1999,gelmanModalDistributional2013}. The derivation in~\eqref{eq:Z_laplace} approximates the posterior $\trg$ as a Gaussian with mean~$\MAP$ and covariance~$\MAPhess^{-1}$. To ensure that the proposal tails are heavy enough, Gelman et al.~\cite{gelmanModalDistributional2013} suggest choosing $q$ to be a $t$~distribution with $\nu=4$ degrees of freedom, mean $\MAP$, and scale matrix $\MAPhess^{-1}$. We call this approach \emph{Laplace importance sampling} (Laplace IS).



However, elliptical proposal distributions are not appropriate for posteriors with challenging geometry that arises from parameter non-identifiability. Consider instead a sequence of $T$ proposal distributions $q_1,\ldots,q_T$, where for each $t=1,\ldots,T$, we have $N_t$ samples $\params^{(t)} = \{\param^{(t)}_n\}_{n=1}^{N_t}$ drawn from $q_t$. The use of multiple proposal distributions provides flexibility for the overall mixture of $q_1,\ldots,q_T$ to more accurately capture the geometry of the posterior $\pi$~\cite{veachOptimallyCombining1995}. Applying the technique of deterministic multiple mixture~\cite{owenSafeEffective2000} leads to the following estimator:
\begin{equation}\label{eq:MIS_estimator}
    \hat{Z} = \frac{1}{\sum_{t=1}^T N_t} \sum_{t=1}^T \sum_{n=1}^{N_t} \frac{\unnormtrg(\param^{(t)}_n)}{q(\param^{(t)}_n)},
\end{equation}
where 
\begin{equation*}
    q(\param) = \frac{1}{\sum_{t=1}^T N_t} \sum_{t=1}^T N_t q_t(\param)
\end{equation*}
is the overall proposal distribution. Moreover, for each $t < T$, we allow the construction of $q_{t+1}$ to depend on the preceding proposals $q_1,\ldots,q_{t}$ and their corresponding samples $\params^{(1)}, \ldots, \params^{(t)}$. In practice, the proposal distributions are taken from some parametric family of distributions, such that adaptation amounts to finding an optimal distribution from the parametric family, where the objective depends on the preceding proposals and corresponding samples. This approach is known as \emph{adaptive multiple importance sampling} (AMIS)~\cite{cornuetAdaptiveMultiple2012, bugalloAdaptiveImportance2017, rafteryEstimatingProjecting2010}; we present the generic AMIS procedure in Algorithm~\ref{alg:generic_amis}. Such a scheme assesses how well the proposals so far approximate the posterior $\trg$ and adapts future proposals accordingly. Although there is no theoretical guarantee that the AMIS estimator is unbiased, in practice, AMIS produces estimators with smaller mean squared errors than non-adaptive importance sampling~\cite{cornuetAdaptiveMultiple2012}.

\begin{algorithm}[t]
\caption{Evidence estimation using adaptive multiple importance sampling}\label{alg:generic_amis}
\begin{algorithmic}[1]
\Require Unnormalized posterior $\unnormtrg$, initial proposal distribution $q_1$, sample size schedule $N_1, \ldots, N_T$
\For{$t = 1, \ldots, T$}
    \State $\params^{(t)} \gets N_t \text{ samples drawn independently from } q_t$
    \If{$t < T$}
    \State Adapt next proposal distribution $q_{t+1}$ using $q_1,\ldots,q_{t}$ and $\params^{(1)}, \ldots, \params^{(t)}$
    \Else  
    \State $q \gets \sum_{t=1}^T \frac{N_t}{N} q_t$
    \State \Return $\hat{Z} \gets \frac{1}{\sum_{t=1}^T N_t} \sum_{t=1}^T \sum_{n=1}^{N_t} \frac{\unnormtrg(\param^{(t)}_n)}{q(\param^{(t)}_n)}$
    \EndIf
\EndFor
\end{algorithmic}
\end{algorithm}

In~\cite{cornuetAdaptiveMultiple2012}, the authors suggest initializing $q_1$ as a logistic distribution, which is an elliptical distribution with finite moments and heavier-than-Gaussian tails. The subsequent proposals $q_2,\ldots,q_T$ are parameterized as Gaussian mixtures:
\begin{equation}\label{eq:gmix}
    q_t(\param) = \sum_{k=1}^K \gmixw_k^{(t)} \phi\!\left(\param; \mu_k^{(t)}, \bm\Sigma_k^{(t)}\right) \quad \text{for } t=2,\ldots,T,
\end{equation}
where $K$ is the number of Gaussian components, $\gmixw_k^{(t)}$ is the weight of component $k$ for proposal~$q_t$, and $\phi(\cdot; \mu, \bm\Sigma)$ denotes the Gaussian density with mean $\mu$ and covariance $\bm\Sigma$. For each $t=1,\ldots,{T-1}$, the mixture parameters $\{(\gmixw_k^{(t+1)}, \mu_k^{(t+1)}, \bm\Sigma_k^{(t+1)})\}_{k=1}^K$ are determined using a weighted Expectation-Maximization (EM) algorithm based on the preceding samples $\params^{(1)}, \ldots, \params^{(t)}$. The weighted EM algorithm seeks to find mixture parameters that minimize $\E[\trg]{-\log q_{t+1}(\param)}$, where the expectation with respect to $\trg$ requires the use of sample weights to account for the fact that the preceding samples are drawn from $q_1,\ldots,q_t$ instead of $\trg$; see Supplementary Materials~S1.1 for details.

We expect AMIS to be more robust against parameter non-identifiability compared to Laplace IS, at the cost of extra computation due to the use of multiple proposals and adaptation. In this work, to compare AMIS with Laplace IS, we implement AMIS by using the Laplace IS proposal distribution as the initial AMIS proposal distribution and adapt the subsequent AMIS proposals using the weighted EM algorithm following~\cite{cornuetAdaptiveMultiple2012}.

\subsection{Bridge sampling}

A potential shortcoming of importance sampling is that the proposal distribution(s) may not adequately cover the posterior $\trg$. A more reliable alternative is to draw samples from the posterior, and post-process these posterior samples to estimate the model evidence using bridge sampling~\cite{mengSimulatingRatios1996a}. Given an importance sampling proposal~$q$ that approximates the posterior $\trg$, bridge sampling iteratively refines a bridge function that mediates between $q$ and $\trg$ to produce an evidence estimate that is less sensitive to the mismatch between $q$ and $\pi$ compared to standard importance sampling. In this work, we implement bridge sampling by drawing posterior samples using the general-purpose No-U-Turn Sampler (NUTS)~\cite{hoffmanNoUturnSampler2014}; see Supplementary Materials~S2 for details. NUTS is a state-of-the-art MCMC method that exploits gradients of $\log\unnormtrg$ to explore the posterior distribution, although repeated gradient evaluation can be computationally expensive for likelihoods that require numerically solving differential equation systems. This can cause bridge sampling to be computationally prohibitive in practical settings where NUTS has to be run for a large number of models with expensive likelihoods. In this work, we use bridge sampling as a gold standard to benchmark other evidence computation methods.

\section{Improving adaptive multiple importance sampling}\label{sec:robust_AMIS}

When implementing AMIS, the choice of the proposal distributions $q_1,\ldots,q_T$ can drastically impact its performance~\cite{cornuetAdaptiveMultiple2012}. For example, if there are parameter regions with significant posterior probability mass that the initial proposal~$q_1$ does not sample from frequently enough, we may require an impractically large number of samples for subsequent proposals to adapt to such regions. When parameter non-identifiability is present, AMIS as described in Section~\ref{sec:IS} can have difficulty in adapting a simple initial proposal towards a posterior with challenging geometry. 

In view of these concerns, we propose a novel AMIS method that features extensions to the initialization and adaptation strategies. These extensions are designed to improve the proposal coverage of posteriors with challenging geometry, and hence robustness against parameter non-identifiability. First, we specify the initial proposal~$q_1$ to be a mixture distribution instead of an elliptical distribution, such that the posterior is better captured from the start to accelerate adaptation. Second, we seek to reduce the variance of the evidence estimate $\hat{Z}$ (see~\eqref{eq:MIS_estimator}) by refining the Gaussian mixture weights of the proposals $q_2, \ldots, q_T$ after each instance of the weighted EM algorithm. 

To construct a mixture distribution as the initial proposal~$q_1$, we run the Pathfinder algorithm~\cite{zhangPathfinderParallel2022} on the unnormalized density $\unnormtrg$. Pathfinder solves the optimization problem of minimizing $-\log \unnormtrg(\param)$ with respect to $\param$ using a multi-start quasi-Newton method and keeps track of Gaussian approximations of the posterior $\trg$ centered at states visited by the optimization path. Although the Pathfinder algorithm was originally designed to efficiently provide initial states for MCMC algorithms, here we use the Gaussian distributions returned by Pathfinder to specify a uniform Gaussian mixture as the initial proposal~$q_1$ to AMIS. We perform multiple runs of Pathfinder from randomly chosen starting points, where each run of Pathfinder returns up to hundreds of Gaussians. When constructing $q_1$, we use only a subset of these Gaussians that are sufficiently dissimilar, such that $q_1$ is likely to broadly cover the parameter regions with significant posterior probability mass. Details about proposal initialization can be found in Supplementary Materials~S1.2.

To motivate variance reduction, recall the importance sampling estimator $\hat{Z}$ in~\eqref{eq:IS_estimator} for a generic proposal~$q$, where samples $\{\param_n\}_{n=1}^N$ are independently drawn from proposal~$q$. In this case, the variance of $\hat{Z}$ can be shown to be
\begin{equation}
    \mathrm{Var}(\hat{Z}) = \frac{1}{N} \Var[q]{\frac{\unnormtrg(\param)}{q(\param)}} = \frac{1}{N}\left(Z \, \E[\trg]{\frac{\unnormtrg(\param)}{q(\param)}}  - Z^2\right), 
\end{equation}
see Supplementary Materials~S1.3 for a derivation. This suggests that a proposal density $q$ that provides estimates of low variance should minimize $\E[\trg]{\unnormtrg(\param)/q(\param)}$. In contrast, at each iteration $t=1,\ldots,T-1$ of AMIS as described in Section~\ref{sec:IS}, the next proposal~$q_{t+1}$ is adapted using the weighted EM algorithm, which instead aims to minimize $ \E[\trg]{-\log q_{t+1}(\param)}$.

At each iteration $t=1,\ldots,T-1$, we perform variance reduction by refining the mixture weights of the next proposal $q_{t+1}$ to minimize $\E[\trg]{\unnormtrg(\param)/q(\param)}$, where $q$ is an overall proposal defined in \eqref{eq:chisq_proposal}. Recall that $q_{t+1}$ is a mixture of $K$ Gaussians with location and scale parameters $\{(\mu_k^{(t+1)}, \bm\Sigma_k^{(t+1)})\}_{k=1}^K$ and mixture weights $\{\gmixw_k^{(t+1)}\}_{k=1}^K$. The overall proposal $q$ represents the distribution of all $N_1+\ldots+N_T$ samples assuming that the $N_{t+1}+\cdots+N_T$ samples yet to be drawn will be drawn from $q_{t+1}$:
\begin{equation}\label{eq:chisq_proposal}
    q(\param) = \sum_{s=1}^t \frac{N_s}{N_1+\cdots+N_T} q_s(\param) + \frac{N_{t+1}+\cdots+N_T}{N_1+\cdots+N_T} \sum_{k=1}^K \gmixw_k^{(t+1)} \phi\!\left(\param; \mu_k^{(t+1)}, \bm\Sigma_k^{(t+1)}\right).
\end{equation}
We minimize $\E[\trg]{\unnormtrg(\param)/q(\param)}$ with respect to the mixture weights $\{\gmixw_k^{(t+1)}\}_{k=1}^K$ only, while the location and scale parameters $\{(\mu_k^{(t+1)}, \bm\Sigma_k^{(t+1)})\}_{k=1}^K$ are fixed at the values obtained from running the weighted EM algorithm beforehand. In Supplementary Materials~S1.3, we show that the resulting minimization problem is convex and provide further information on the computational details.

\section{Results}

In this section, we evaluate evidence estimation methods for ordinary differential equation (ODE) models where parameter non-identifiability is present. Using bridge sampling as a gold standard, we compare the model evidence and model posterior distributions obtained with BIC, Laplace IS, AMIS as described in Section~\ref{sec:IS}---which we call standard AMIS, and the new implementation of AMIS described in Section~\ref{sec:robust_AMIS}---which we call robust AMIS. We omit the Laplace approximation in our comparisons as it is equivalent to performing Laplace IS with a single deterministic ``sample'', namely the MAP estimate. We use the \emph{total variation distance} (TVD) to compare model posterior distributions obtained with different methods. Given model posterior distributions $\hat{p}_1(\model\vert\data)$ and $\hat{p}_2(\model\vert\data)$ obtained with two different methods, the TVD between $\hat{p}_1$ and $\hat{p}_2$ is defined as
\begin{equation*}
    \mathrm{TVD}(\hat{p}_1,\hat{p}_2) = \frac{1}{2}\sum_{\model\in\modelspace} \lvert \hat{p}_1(\model\vert\data) - \hat{p}_2(\model\vert\data) \rvert.
\end{equation*}
The TVD is bounded between zero and unity, and is equivalent to the maximum possible discrepancy in the posterior probability of any model subset, i.e.
\begin{equation*}
    \mathrm{TVD}(\hat{p}_1,\hat{p}_2) = \max_{\modelspace^*\subseteq\modelspace}  \left\vert \sum_{\model\in\modelspace^*} \hat{p}_1(\model\vert\data) - \sum_{\model\in\modelspace^*} \hat{p}_2(\model\vert\data) \right\vert.
\end{equation*}

For the importance sampling methods, we draw a total of $N=10^6$ proposal samples and apply Pareto smoothing~\cite{JMLR:v25:19-556} for improved estimator stability. For both AMIS methods, we use a sample size schedule $N_1, \ldots, N_T$ of $T=16$ iterations, where the cumulative sums $\{\sum_{s=1}^t N_s\}_{t=1}^T$ form a geometric sequence from $10^4$ to $10^6$. The use of increasing sample sizes follows the recommendation in~\cite{marinConsistencyAdaptive2019} that the majority of the computational budget should be allocated to the final iterations of adaptive importance sampling algorithms. Algorithmic settings for all methods are reported in Supplementary Table~S1.

A key diagnostic metric for importance sampling schemes is the \emph{effective sample size} (ESS). For an importance sampling estimator of the form $\hat{Z} = N^{-1} \sum_{n=1}^N w(\param_n)$, e.g., $w(\param) = \unnormtrg(\param)/q(\param)$ for \eqref{eq:IS_estimator}, the ESS is given by
\begin{equation}\label{eq:ess}
    \mathrm{ESS} = \frac{\left( \sum_{n=1}^N w(\param_n) \right)^2}{\sum_{n=1}^N w(\param_n)^2}.
\end{equation}
The quantity $w(\param)$ is known as the \emph{importance weight} of $\param$. In the ideal case where the proposal is exactly the posterior, $w(\cdot)$ is constant, leading to the maximum possible ESS of $N$. Conversely, a small ESS is indicative of large outliers in the importance weights, which correspond to parameter values $\param$ that are underrepresented by the proposal compared to the posterior.

Note that all parameters of the models that we consider are constrained to be positive, and do not vary over time. However, the importance sampling schemes are designed for unconstrained parameter spaces. As such, we apply a $\log_{10}$ transformation on all model parameters and impose independent, wide normal priors on the log-transformed parameters $\param$ (see Supplementary Materials~S3). We assume throughout that all candidate models are \textit{a priori} equally likely, i.e.~the prior model distribution is $p(\model)=1/M$ for each model $\model\in\modelspace$. A uniform prior over models can be viewed as a neutral choice~\cite{hoetingBayesianModel1999}; see \cite{hoetingBayesianModel1999,scottBayesEmpiricalBayes2010,oatesCausalNetwork2014} for alternatives that incorporate domain knowledge into the model prior.

\subsection{Coral re-growth models}\label{sec:coral}

We apply Bayesian model selection to $M=3$ sigmoid growth models considered in~\cite{simpsonParameterIdentifiability2022} for modeling the re-growth of hard coral covers, namely the logistic, Gompertz, and Richards' models. Each model describes the growth of a single coral population $C$ over time $t$ with an ODE:
\begin{align}
    \frac{\dd{C}}{\dd{t}} &= rC\left(1-\frac{C}{K}\right), \tag{Logistic} \\
    \frac{\dd{C}}{\dd{t}} &= rC\log\!\left(\frac{K}{C}\right), \tag{Gompertz} \\
    \frac{\dd{C}}{\dd{t}} &= rC\left[1-\left(\frac{C}{K}\right)^\beta\right]. \tag{Richards'}    
\end{align}
The model parameters consist of the ODE parameters---such as $(r,K,\beta)$ for the Richards' model, the initial coral population $C(0)$, and an additive noise standard deviation $\sigma$. Details pertaining to the model parameters and noise distribution can be found in Supplementary Materials~S3.1. The logistic model is a special case of the Richards' model where $\beta=1$, the Gompertz model is a limiting case (up to reparameterization) of the Richards' model as $\beta\rightarrow 0^+$~\cite{simpsonParameterIdentifiability2022}. Note that for all three models, exact ODE solutions are analytically available, so the likelihood is computationally cheap to evaluate. This is not typical of most mathematical biology applications, which generally require numerical schemes for solving ODEs. Nevertheless, the availability of exact solutions gives a convenient setup to repeatedly test evidence estimation.

\begin{figure}[t]
\centering
\includegraphics[width = 0.99\textwidth]{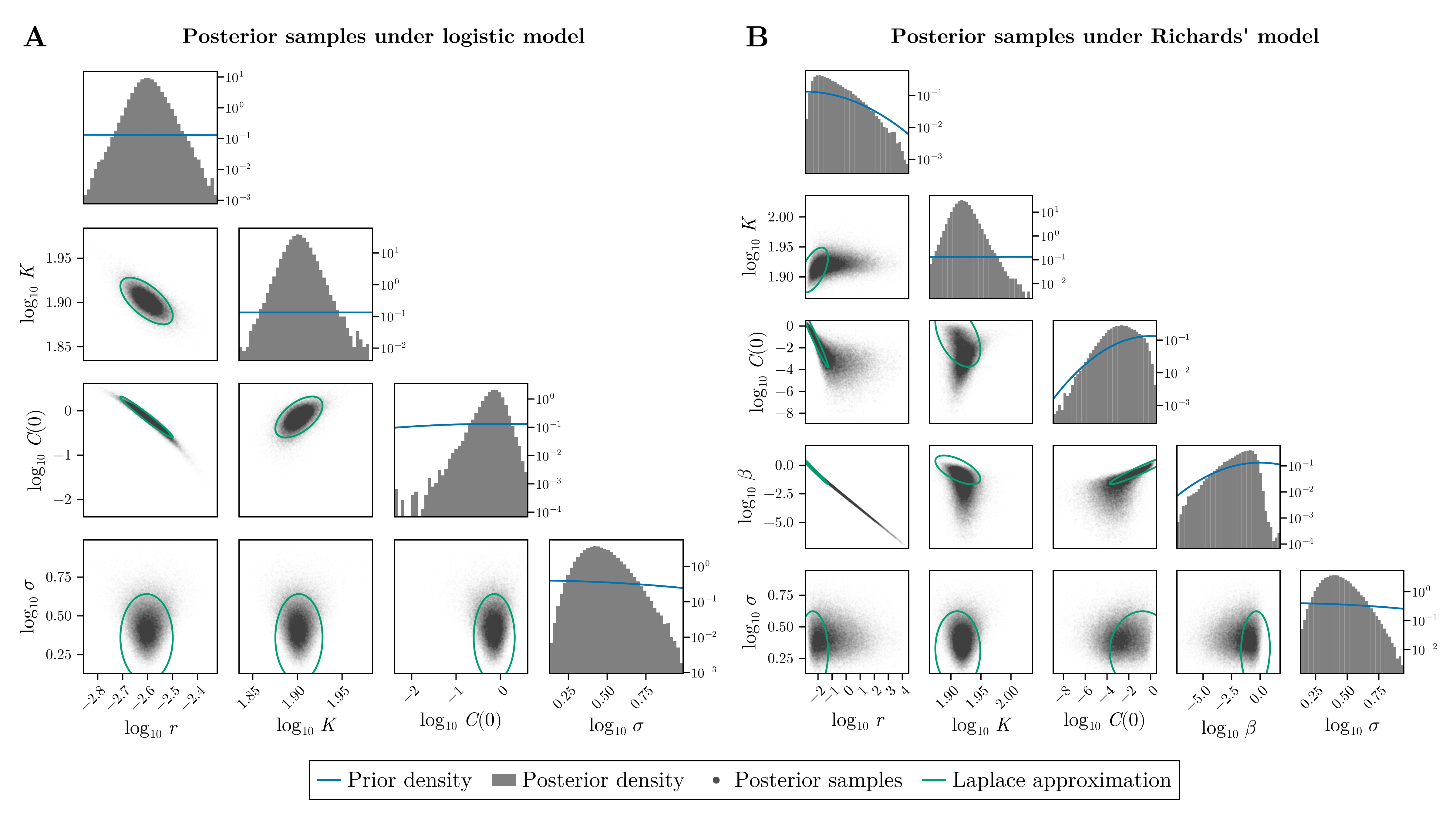}
\caption{Posterior samples obtained from one NUTS run consisting of five independent chains, under (A)~the logistic model and (B)~the Richards' model; the posterior under the Gompertz model is shown in Supplementary Figure~S2. Plots along the main diagonals show univariate marginals of the posterior (histogram) along with the prior (blue lines), where density is shown on a log scale. Plots off the main diagonals show bivariate marginals of the posterior (gray points) along with the 95\% probability region of the Gaussian approximation centered at the MAP estimate (green ellipses).}
\label{fig:logistic_richards}
\end{figure}

We assess the practical identifiability of these three models given the data reported in the main text of~\cite{simpsonParameterIdentifiability2022}, which consists of $11$ time-course measurements of hard coral cover over a decade (plotted in Supplementary Figure~S1). In Bayesian inference, practical non-identifiability is characterized by the similarity of the parameter prior and posterior densities: if the data is uninformative in constraining a parameter over some parameter range, the likelihood varies slowly over that range, leading to a posterior density that is approximately proportional to the prior density. We use NUTS to draw samples from the posterior under each of the three models. For each model, we run five independent chains of $2\times 10^4$ samples each after discarding $10^3$ burn-in samples, resulting in a total of $10^5$ posterior samples. Figure~\ref{fig:logistic_richards}A shows peaked posterior densities under the logistic model, along with prior densities that are near constant over the effective domain of the posterior. In Figure~\ref{fig:logistic_richards}B, we see that the posterior and prior densities are similar under the Richards' model over the effective domain of the posterior for the parameters $r, C(0), \beta$, concurring with the parameters that were found to be non-identifiable in~\cite{simpsonParameterIdentifiability2022}. For the Gompertz model, we find that the $C(0)$ parameter is non-identifiable (Supplementary Figure~S2).

We observe that where parameters are non-identifiable, the corresponding posteriors exhibit stronger non-Gaussianity. In each off-diagonal plot of Figure~\ref{fig:logistic_richards}, we visualize the Laplace approximation with the 95\%~probability region of the Gaussian with mean $\MAP$ and covariance $\MAPhess^{-1}$ (see Section~\ref{sec:bic}) under the corresponding model. For the Richards' model, we see that the 95\%~probability region misses large portions of the bivariate posterior marginals involving the poorly identified parameters $r, C(0), \beta$. Moreover, some missed portions do not align with the dominant eigendirections of the covariance $\MAPhess^{-1}$, e.g., the bivariate marginal of $(r, C(0))$, indicating that the use of any single elliptical distribution is unlikely to cover the posterior adequately. Thus, we expect BIC and Laplace IS to underestimate the evidence of the Richards' model.

\begin{figure}[t]
\centering
\includegraphics[width = 0.99\textwidth]{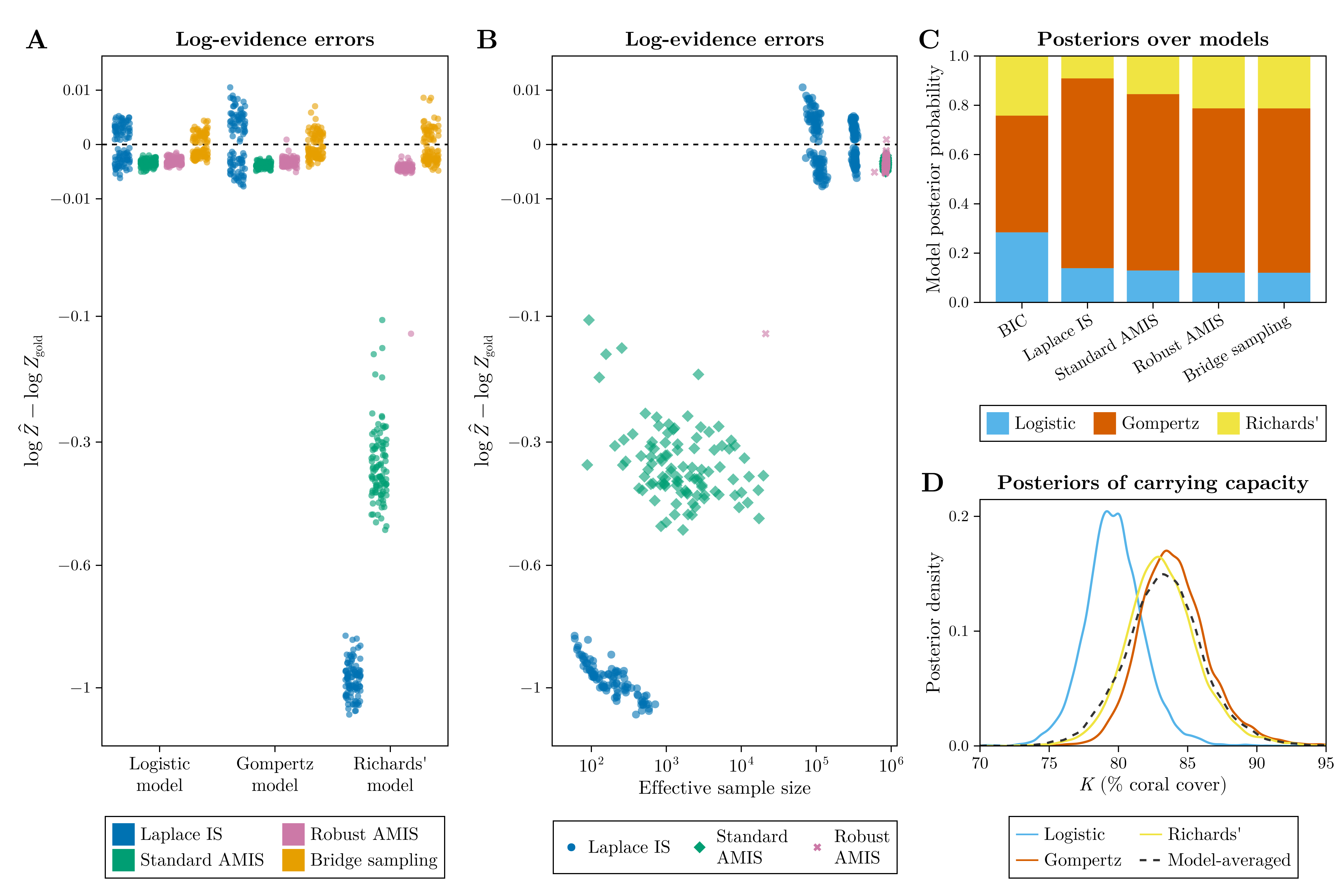}
\caption{Bayesian model selection results for the coral re-growth example obtained with each method. (A)~Estimation errors of log-evidence (over 100 runs per method) compared to the gold-standard value $Z_\text{gold}$, where the vertical axis follows a square root scale. (B)~Log-evidence errors from panel A plotted against the corresponding ESS; models are not visually distinguished here. (C)~Model posterior probabilities obtained using one representative run of each method. (D)~Single-model and model-averaged posterior distributions of the carrying capacity $K$ obtained with a representative run of robust AMIS.}
\label{fig:comparison_coral}
\end{figure}

We repeatedly estimate the evidence of each model 100 times using Laplace IS, standard AMIS, robust AMIS, and bridge sampling. BIC is omitted in this comparison as it is a deterministic approximation with an $\bigO{1}$ error. For each run of bridge sampling, we use a separate run of NUTS, such that the posterior samples used are independent across bridge sampling runs. All NUTS runs returned a potential scale reduction factor~\cite{vehtariRanknormalizationFolding2021} of $\hat{R}<1.01$, indicating that the chains are likely to have mixed well. To evaluate the accuracy of evidence estimates $\hat{Z}$ obtained with each method, we take the average of the evidence estimates obtained over 100 runs of bridge sampling to be the gold-standard value $Z_\text{gold}$ of the model evidence. In Figure~\ref{fig:comparison_coral}A, we see that Laplace IS and standard AMIS underestimate the evidence of the Richards' model. This underestimation is explained by the fact that the non-identifiability of the Richards' model leads to a Laplace approximation that does not cover the posterior adequately. Thus, parameter regions of significant posterior probability mass are ignored by Laplace IS and standard AMIS, which use the Laplace approximation as the only proposal distribution and the initial proposal distribution, respectively. Comparing errors between the logistic and Gompertz models, we see that the variance of the Laplace IS estimator increases, but this is not the case for standard AMIS. This is most likely explained by the poor identifiability of parameter $C(0)$ for the Gompertz model, leading to a non-Gaussian posterior; see the bivariate posterior marginal of $(r, C(0))$ in Supplementary Figure~S2. As the posterior regions missed by the Laplace approximation are not large, standard AMIS successfully adapts to the posterior so its estimator achieves a small variance for the Gompertz model.

The robust AMIS estimates are significantly less biased than standard AMIS estimates for the Richards' model (Figure~\ref{fig:comparison_coral}A); noting that the vertical axis follows a square root scale (for visual clarity of near-zero values). The estimated bias and standard deviation of all Monte Carlo estimates are reported in Supplementary Table~S2. All but one robust AMIS run result in a log-evidence error of magnitude $<\!10^{-2}$, where the outlying robust AMIS run has a much smaller ESS than the other robust AMIS runs (Figure~\ref{fig:comparison_coral}B). This particular run is likely due to a poor initialization of the proposal $q_1$ that failed to adequately cover the posterior $\trg$ during adaptation. Overall, robust AMIS is capable of producing accurate evidence estimates for the Richards' model where standard AMIS fails. The magnitude of bridge sampling log-evidence errors is $<\!10^{-2}$ across all models and runs, instead of degrading as non-identifiability becomes more severe. This is because NUTS automatically computes longer leapfrog trajectories to explore more challenging posterior geometries that come with non-identifiability; NUTS takes, on average, 1 minute, 1 minute, and 3 minutes to run for the logistic, Gompertz, and Richards' models, respectively. All computational times in this work are reported for a single Intel(R) Xeon(R) Gold 6448H core.

Figure~\ref{fig:comparison_coral}C illustrates the downstream impact of evidence estimation on computing model posterior probabilities (results shown only for a representative run from all repetitions). The posterior probabilities obtained with robust AMIS and bridge sampling are virtually identical. Laplace IS and standard AMIS underestimate the posterior probability of the Richards' model and overestimate the posterior probability of the other two models. The posterior probability of the logistic model is overestimated by BIC. A likely explanation is that BIC underestimates the model evidence of the Gompertz and Richards' models due to parameter non-identifiability. This demonstrates the danger of ranking models using the BIC when non-identifiability is present.

To illustrate BMA, we plot in Figure~\ref{fig:comparison_coral}D the single-model and model-averaged posterior distributions of the carrying capacity $K$, which can be interpreted as the long-term prediction of the coral population $C(t)$ as $t\rightarrow \infty$. The single-model posteriors are constructed using the samples and importance weights returned by robust AMIS; the model-averaged posterior is a combination of the single-model posteriors weighted according to the model posterior probabilities obtained with robust AMIS. We note that BMA applies similarly to prediction trajectories that can be computed as a function of the parameters and the model (see Section~\ref{sec:model_selection}).

\subsection{Insect life-stage models}\label{sec:insect}

To evaluate evidence estimation methods in the context of larger model selection problems, we conduct a simulation study involving $M=64$ life-stage population models of a hypothetical insect with egg, larva, and adult stages. We denote the population size of these stages as $E$, $L$, and $A$, respectively. We assume that individuals at each stage may die at a first-order rate due to processes unrelated to the population, and/or at a second-order rate due to competition between insects of the same life stage. In the case where both death mechanisms are present for all stages, we model the population dynamics with the following ODE system:
\begin{align*}
    \frac{\dd{E}}{\dd{t}} &= \rho A - \lambda_{EL} E - \delta_E E - \frac{1}{2}\kappa_E E^2, \\
    \frac{\dd{L}}{\dd{t}} &= \lambda_{EL} E - \lambda_{LA} L - \delta_L L - \frac{1}{2}\kappa_L L^2, \\
    \frac{\dd{A}}{\dd{t}} &= \lambda_{LA} L - \delta_A A - \frac{1}{2}\kappa_A A^2,
\end{align*}
where $\rho$ is the reproduction rate, $\lambda_{EL},\lambda_{LA}$ are inverse durations of the egg and larva stages, $\delta_E,\delta_L,\delta_A$ are first-order death rates of each stage, and $\kappa_E,\kappa_L,\kappa_A$ are second-order death rates of each stage; see Supplementary Materials~S3.2 for an explanation of the factor of $1/2$ appearing in the ODE system. We consider the reproduction and stage transitions to be core mechanisms, and ask the model selection question of which of the six death mechanisms are present. Specifically, by exhaustively setting each possible subset of $\{\delta_E,\delta_L,\delta_A,\kappa_E,\kappa_L,\kappa_A\}$ to be zero, we obtain a model set~$\modelspace$ of $M=64$ models. We generate 44 synthetic datasets, each based on a different model where all populations tend to a positive equilibrium. The remaining 20 models result in population extinction or unbounded population growth. The datasets are plotted in Supplementary Figure~S3; details about dataset generation, parameter priors and likelihoods are presented in Supplementary Materials~S3.2.

We consider 44 model selection problems---one for each dataset---over the model set~$\modelspace$ of 64 models. For each of the $44\times 64=2816$ dataset-model pairs, we estimate the evidence of the model given the dataset using each method. We do not repeat runs as in Section~\ref{sec:coral}, as this would be computationally prohibitive with the larger model set and cost of solving the ODE system numerically. To perform bridge sampling for each dataset-model pair, we run five independent NUTS chains of $3000$ samples each after discarding $1000$ burn-in samples, resulting in a total of $15000$ posterior samples. The MCMC convergence diagnostics are not ideal: 89\% of NUTS runs returned a potential scale reduction factor~\cite{vehtariRanknormalizationFolding2021} of $\hat{R}<1.01$, while the largest value of $\hat{R}$ encountered across all runs was 1.047. Nevertheless, we consider the model selection results obtained with bridge sampling to be the gold standard to benchmark the other methods against.

\begin{figure}[t]
\centering
\includegraphics[width = \textwidth]{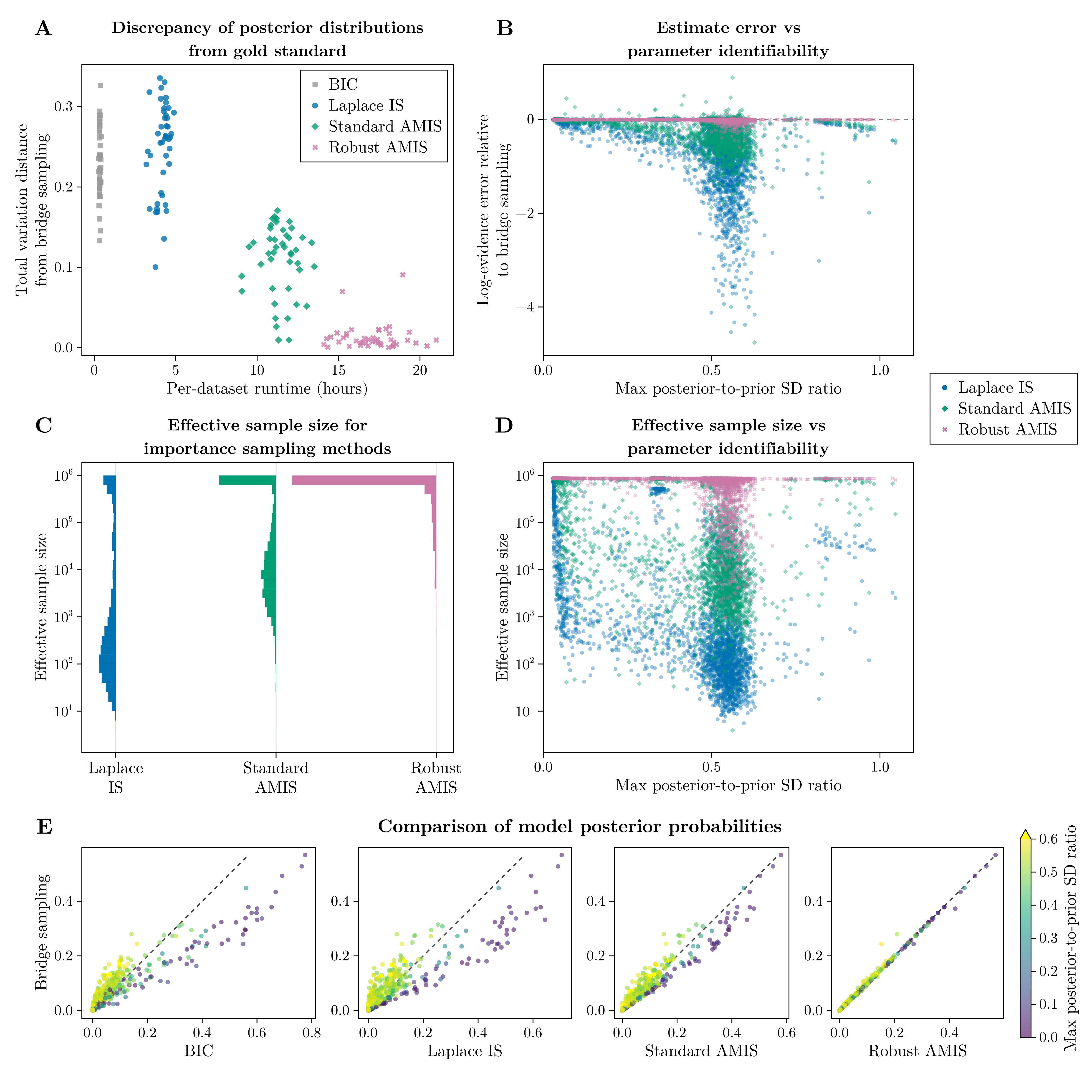}
\caption{Comparison of evidence estimation methods for the insect life-stage example obtained with each method. (A)~Total variation distance between model posterior distributions obtained with bridge sampling and every other method for each dataset. (B)~Difference in log-evidence estimates obtained with bridge sampling and importance sampling methods against a proxy for parameter non-identifiability for each dataset-model pair. (C)~Effective sample sizes returned by importance sampling methods for each dataset-model pair. (D)~Effective sample sizes against a proxy for parameter non-identifiability for each dataset-model pair. (E)~Model posterior probabilities obtained with bridge sampling, compared to those obtained with every other method for each dataset-model pair. Darker colors correspond to cases where parameters are well-identified.}
\label{fig:comparison_insect}
\end{figure}

The average computational time taken per dataset to run NUTS for 64 models is 375 hours; the post-processing of posterior samples involved in bridge sampling takes on average 5.4 hours per dataset. The per-dataset runtime of all other methods is substantially lower than that of bridge sampling (Figure~\ref{fig:comparison_insect}A). Compared to BIC, the main cost of Laplace IS comes from having to numerically solve an ODE system $N=10^6$ times for each model, a cost that is common to the two AMIS implementations and the post-processing of posterior samples by bridge sampling. The average runtime of standard AMIS and robust AMIS are of the same order of magnitude. 

For each model selection problem, we evaluate BIC and the importance sampling methods using the TVD between the model posterior returned by bridge sampling (gold standard) and the model posterior returned by each other method (Figure~\ref{fig:comparison_insect}A). BIC and Laplace IS are inaccurate in reproducing the gold-standard model posterior distribution, standard AMIS achieves an average TVD of 0.11, while robust AMIS improves this to 0.013. Robust AMIS produces more accurate model selection results than standard AMIS within a similar computational budget, while taking substantially less time than bridge sampling. We note that the improved accuracy from Laplace IS to standard AMIS to robust AMIS correlates with increasing ESS (Figure~\ref{fig:comparison_insect}C). 

The error in computing a model posterior distribution consists of errors in evidence estimation. Across the 2816 dataset-model pairs, the proportion of log-evidence estimates that are within 0.1 of the corresponding bridge sampling estimate is 38\%, 55\%, 98\% for Laplace IS, standard AMIS, robust AMIS, respectively. To investigate whether the accuracy of evidence estimation correlates with practical identifiability, we quantify identifiability via the maximum ratio of the posterior parameter standard deviations (SDs) to the prior parameter SDs, where the maximum is taken over parameters. Larger posterior-to-prior SD ratios correspond to parameters that are not well-constrained by the data. From Figures~\ref{fig:comparison_insect}B and~\ref{fig:comparison_insect}D, we observe that as the maximum posterior-to-prior SD ratio increases, that is, as non-identifiability worsens, the evidence estimation tends to be more biased, and ESS decreases. This effect is strongest for Laplace IS, followed by standard AMIS, demonstrating that parameter non-identifiability presents a challenge for estimating model evidence accurately. About half of the dataset-model pairs have maximum posterior-to-prior SD ratios between 0.5 and 0.6, which likely correspond to parameters that are only one-sided identifiable; see for example parameters $\delta_E,\delta_L,\kappa_E,\kappa_A$ in Supplementary Figure~S4. For reference, the standard deviation of a standard half-normal distribution is 0.6. The prevalence of parameters that are only one-sided identifiable aligns with analyses of parameter identifiability in mathematical biology literature~\cite{raueStructuralPractical2009,browningIdentifiabilityAnalysis2020,browningGeometricAnalysis2023}. We find that BIC, Laplace IS, and standard AMIS underestimate the model posterior probabilities for dataset-model pairs with larger maximum posterior-to-prior SD ratios (Figure~\ref{fig:comparison_insect}E), i.e. cases where parameters are poorly identified. Since model posterior probabilities must sum to unity, these methods overestimate the posterior probabilities of models whose parameters are well-identified (small maximum posterior-to-prior SD ratio).

\begin{figure}[t]
\centering
\includegraphics[width = 0.9\textwidth]{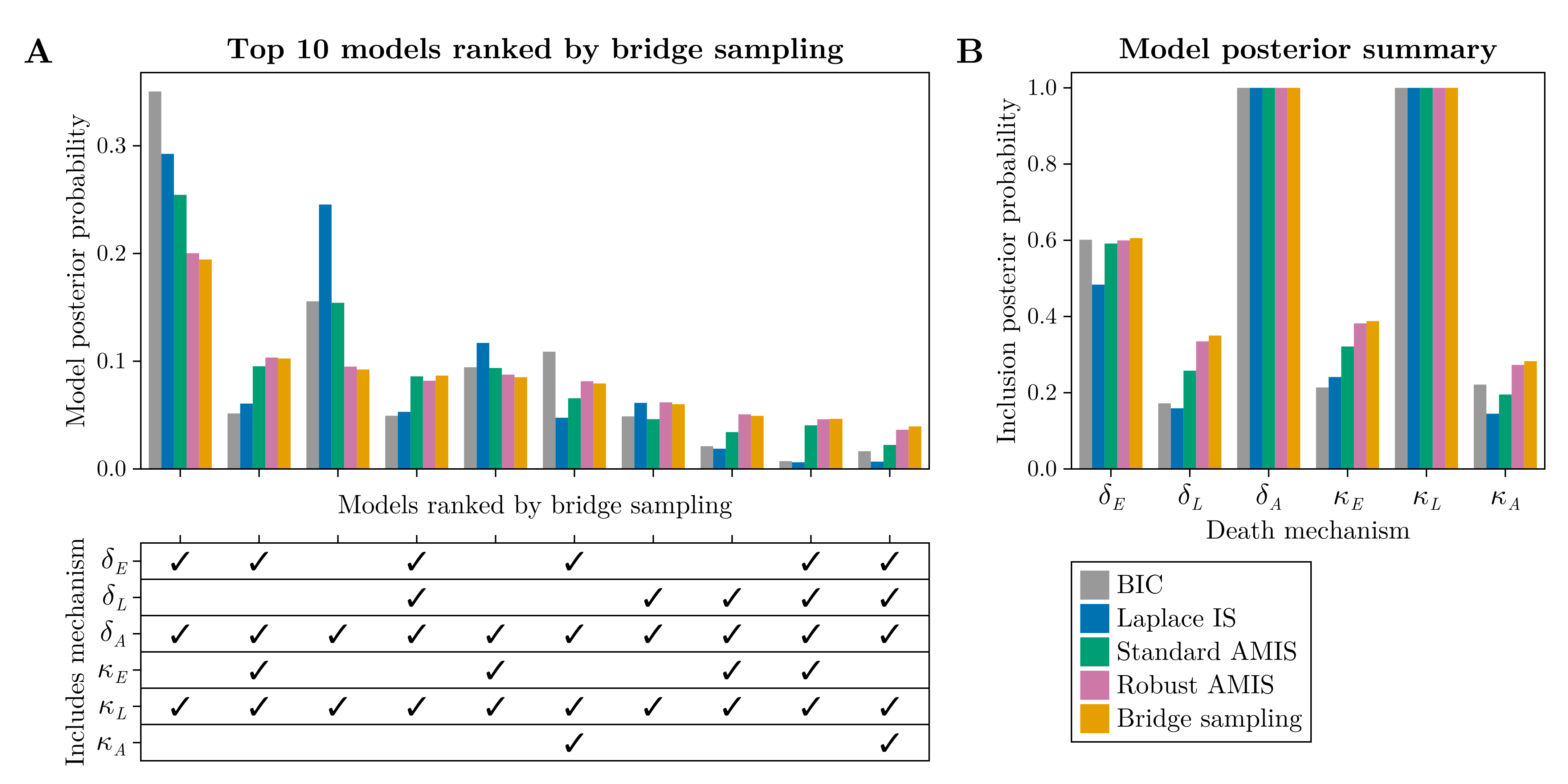}
\caption{Model selection summaries obtained with each method for one of the insect life-stage datasets. (A)~Model posterior probabilities of the top ten models ranked by posterior probabilities obtained with bridge sampling, with the bottom panel indicating the death mechanisms included in each model. (B)~Model-averaged probabilities for the inclusion of each possible death mechanism in the model.}
\label{fig:modelposts_insect}
\end{figure}

We illustrate the model selection results obtained with each method for one of the datasets in Figure~\ref{fig:modelposts_insect}. The dataset is generated using the model that includes only the death mechanisms that correspond to the parameters $\delta_E,\delta_L,\delta_A,\kappa_L$. Figure~\ref{fig:modelposts_insect}A shows the model posterior probabilities of the top ten models ranked by their posterior probabilities obtained with bridge sampling. The posterior probabilities obtained with robust AMIS and bridge sampling are in close agreement, with robust AMIS being 18~times faster than bridge sampling for this example. The discrepancy between standard AMIS and bridge sampling largely comes from standard AMIS overestimating the posterior probability of the models ranked first and third by bridge sampling, which have fewer death mechanisms included. BMA allows us to compute the posterior probabilities of each death mechanism being included in the model (Figure~\ref{fig:modelposts_insect}B). The posterior probabilities of including the mechanisms corresponding to the parameters $\delta_A$ and $\kappa_L$ are both close to unity, indicating that the data strongly supports the presence of the these two death mechanisms. BIC, Laplace IS, and standard AMIS underestimate some of the probabilities in Figure~\ref{fig:modelposts_insect}B compared to the gold standard results obtained with bridge sampling. A possible explanation is that models with more parameters are more likely to be over-parameterized, and thus more susceptible to parameter non-identifiability, which correlates with underestimated model posterior probabilities (Figure~\ref{fig:comparison_insect}E).

\section{Discussion}

Despite the widespread use of modeling in mathematical biology, there is little practical guidance in the literature on the choice of model selection methods when parameter non-identifiability is present. The use of information criteria is pervasive~\cite{kirkModelSelection2013,tredennickPracticalGuide2021,zhangInformationCriteria2023}, though most information criteria do not adequately account for parameter uncertainty. We find that evidence estimation methods that heavily rely on the posterior being elliptical---such as BIC and Laplace IS---are unreliable when parameters are non-identifiable. When non-identifiability is present, regions of significant posterior probability can be missed by elliptical approximations, resulting in model evidence being underestimated (Figures~\ref{fig:comparison_coral} and~\ref{fig:comparison_insect}). In the same vein, importance sampling schemes can produce inaccurate results when the proposal distribution does not cover the posterior distribution sufficiently. Although we have only assessed examples exhibiting practical non-identifiability, we  expect these findings to hold also for cases exhibiting structural non-identifiability, as the concerns regarding highly non-Gaussian posteriors (Section~\ref{sec:bic}) hold for both practical and structural non-identifiability.

In this work, we propose robust AMIS, which we demonstrate to be more robust than standard AMIS against parameter non-identifiability. Robust AMIS replaces the elliptical initial proposal used in standard AMIS with a mixture proposal, such that the initial proposal has better coverage of the posterior. Moreover, standard AMIS adapts Gaussian mixture proposals by minimizing $\E[\trg]{-\log q(\param)}$, which we refine in robust AMIS by minimizing $\E[\trg]{\unnormtrg(\param)/q(\param)}$ with respect to Gaussian mixture weights to reduce estimation variance. Since $1/q(\param)$ diverges faster than $-\log q(\param)$ as $q(\param)\rightarrow0^+$, it is possible that adaptation in standard AMIS does not sufficiently prioritize posterior regions that are assigned very low proposal densities. The minimization of $\E[\trg]{-\log q(\param)}$ and $\E[\trg]{\unnormtrg(\param)/q(\param)}$ is equivalent to minimizing the Kullback-Leibler divergence and the $\chi^2$ divergence, respectively; see \cite{mullerNeuralImportance2019} for further discussion about the use of these two divergences in neural importance sampling. Neural importance sampling uses normalizing flows as proposal distributions, which are more expressive than Gaussian mixtures, but are more computationally demanding. It is unclear whether normalizing flow proposals can outperform Gaussian mixture proposals in terms of evidence estimation efficiency for high-dimensional posteriors.

We recommend practitioners who routinely use the BIC for model selection to check for identifiability issues before trusting BIC results. Where computational costs are not a concern, evidence estimation methods that reliably explore the parameter posterior distribution are more accurate, e.g., bridge sampling with MCMC posterior samples. Methods such as Sequential Monte Carlo~\cite{delmoralSequentialMonte2006,toniApproximateBayesian2008} and nested sampling~\cite{skillingNestedSampling2006}, although not tested in this work, are likely to also be considered computationally costly but accurate, as long as the posterior sampling is robust against challenging geometries~\cite{snyderObstaclesHighdimensional2008,beskosErrorBounds2014,dittmannNotesPractical2024}. These methods can be computationally prohibitive if the model space consists of a large number of models, in which case, robust AMIS can produce evidence estimates comparable to those produced by bridge sampling at a substantially lower computational cost. However, if the number of models is combinatorially large, we would instead require methods that sample over the model space, such as reversible-jump MCMC~\cite{greenReversibleJump1995,galagaliExploitingNetwork2019}.

Occasionally, robust AMIS returns slightly inaccurate results, which can be diagnosed by checking the ESS as defined in \eqref{eq:ess}. A small ESS is indicative of large outliers in the importance weights, which correspond to samples from regions that are insufficiently covered by the proposal when compared to the posterior. We note that ESS is not a perfect diagnostic: if there are poorly-covered regions that an importance sampling scheme never samples from, the ESS returned can be misleadingly large despite the use of a proposal that does not adequately cover the posterior. This can occur for robust AMIS if the posterior is multimodal and Pathfinder fails to find all posterior modes when constructing the initial proposal; see \cite{zhangPathfinderParallel2022} for further discussion on the use of Pathfinder for multimodal distributions. Although we did not attempt to improve the runs with small ESS in this work, we anticipate that such runs can be salvaged by rerunning robust AMIS with increased initialization effort; using more components in the initial mixture proposal and performing more runs of Pathfinder can broaden the proposal coverage of the posterior before adaptation. Alternatively, given that not many robust AMIS runs suffer from small ESS, one can fall back to more computationally intensive methods that estimate evidence accurately. We consider the development of diagnostics and resolutions for robust AMIS runs with small ESS to be future work. Nevertheless, the TVD between the model posterior distributions returned by robust AMIS and bridge sampling is less than 0.1 across all examples, which is better than the average TVD of any other method in the simulation study (Section~\ref{sec:insect}). Although the examples in this work feature model sets that have a nested structure (up to reparameterization), we note that Bayesian model selection is not limited to nested models.

Parameter non-identifiability is a pervasive feature of mathematical models in biology, and its consequences for model selection are easily overlooked when simple evidence computation methods are used. As demonstrated in this work, methods that rely on the Laplace approximation---including BIC---tend to underestimate the evidence of non-identifiable models, systematically biasing model selection in favor of models with well-identified parameters (Figure~\ref{fig:comparison_insect}E). Although there may be scenarios where one \textit{a priori} prefers to select identifiable models~\cite{wangBayesianIdentifiability2026}, we note that non-identifiable models can still have predictive power~\cite{grabowskiPredictivePower2023}, as long as one correctly accounts for the uncertainty of non-identifiable parameters. Robust AMIS is a novel method that addresses this by providing evidence estimates that are accurate even when parameters are non-identifiable, while remaining more computationally efficient than methods that require full posterior sampling, such as bridge sampling with NUTS. We therefore advocate robust AMIS as a practical method for evidence estimation in mathematical biology settings where parameter non-identifiability is anticipated.

\section*{Data availability}

The code and data used in this work is available at \url{https://github.com/ysfoo/crn-model-selection}.

\AtNextBibliography{\small}
\emergencystretch 2em
\printbibliography[heading=bibintoc]

\end{refsection}

\clearpage

\begin{center}
\makeatletter
{\LARGE Supplementary Materials: \@title}\\[0.5em]
{\large \AB@authlist\par\AB@affillist}
\makeatother
\end{center}




\setcounter{section}{0}
\setcounter{equation}{0}
\setcounter{figure}{0}
\setcounter{table}{0}
\setcounter{algorithm}{0}
\makeatletter
\renewcommand{\thesection}{S\arabic{section}}
\renewcommand{\theequation}{S\arabic{equation}}
\renewcommand{\thefigure}{S\arabic{figure}}
\renewcommand{\thetable}{S\arabic{table}}
\renewcommand{\thealgorithm}{S\arabic{algorithm}}

\startcontents[supplementary]
\printcontents[supplementary]{}{1}{\section*{Contents}\vspace{0.5em}}
\begin{refsection}

\section{Details for adaptive multiple importance sampling (AMIS)}

In this section, we describe how the proposal distributions $q_1,\ldots,q_T$ are constructed in standard AMIS and robust AMIS. For each $t=1,\ldots,T$, we draw $N_t$ samples $\params^{(t)} = \{\param^{(t)}_n\}_{n=1}^{N_t}$ from proposal~$q_t$. In standard AMIS, the initial proposal $q_1$ is the Laplace importance sampling (Laplace IS) proposal distribution; in robust AMIS, the initial proposal $q_1$ is a Gaussian mixture (Section~\ref{sec:pathfinder}). For both methods, at each AMIS iteration~$t<T$, we construct a Gaussian mixture proposal~$q_{t+1}$ based on all preceding samples $\params^{(1)},\ldots,\params^{(t)}$. We first compute a placeholder Gaussian mixture $q^\text{pl}_{t+1}$ with $K^\text{pl}$ components using a weighted expectation-maximization (EM) algorithm (Section~\ref{sec:wem}), then remove all components with mixture weights smaller than $10^{-4}$ and rescale the remaining mixture weights (so that they sum to unity) to construct the actual Gaussian mixture $q_{t+1}$ which we draw samples $\params^{(t+1)}$ from. Components with negligible weight are likely to be degenerate, so their removal reduces computational cost with little impact on the proposal distribution. In the case of robust AMIS, the placeholder Gaussian mixture $q^\text{pl}_{t+1}$ is adjusted to reduce estimation variance (Section~\ref{sec:var_reduction}) before removing negligible components.

\subsection{Weighted EM algorithm}\label{sec:wem}

To understand why a weighted EM algorithm is required, we first review the standard EM algorithm for fitting a Gaussian mixture $q$ to data $x_1,\ldots,x_N$ by minimizing 
\begin{equation}\label{eq:std_EM_obj}
    - \frac{1}{N}\sum_{n=1}^N \log q(x_n),
\end{equation}
with respect to the Gaussian mixture parameters, which is equivalent to performing maximum likelihood estimation. The expression in \eqref{eq:std_EM_obj} is an estimate of $\E[\trg_x]{-\log q(x)}$, where $\trg_x$ is the data-generating distribution for the data $x_1,\ldots,x_N$. We write the density of a Gaussian mixture of $K$ components as
\begin{equation*}
    q(x) = \sum_{k=1}^K \gmixw_k \phi\!\left(x; \mu_k, \bm\Sigma_k\right),
\end{equation*}
where $\gmixw_k$ is the weight of component $k$, and $\phi(\cdot; \mu, \bm\Sigma)$ denotes the Gaussian density with mean $\mu$ and covariance $\bm\Sigma$. The standard EM algorithm repeats, for each $k=1,\ldots,K$, the updates
\begin{align*}
    \resp_{n,k} &\leftarrow \frac{\gmixw_k \phi\!\left(x_n; \mu_k, \bm\Sigma_k\right)}{\sum_{j=1}^K \gmixw_j \phi\!\left(x_n; \mu_j, \bm\Sigma_j\right)} \qquad \text{for } n=1,\ldots,N, \\
    \gmixw_k &\leftarrow \frac{\sum_{n=1}^N \resp_{n,k}}{\sum_{n=1}^N \sum_{j=1}^{K} \resp_{n,j}} , \\ 
    \mu_k &\leftarrow \frac{\sum_{n=1}^N \resp_{n,k} x_n}{\sum_{n=1}^N \resp_{n,k}} , \\ 
    \bm\Sigma_k &\leftarrow \frac{\sum_{n=1}^N \resp_{n,k} (x_n-\mu_k)(x_n-\mu_k)^\trp}{\sum_{n=1}^N \resp_{n,k}} ,
\end{align*}
until convergence or when some maximum number of EM iterations is reached, where the $\resp_{n,k}$ are known as \emph{responsibilities}~\cite{bishopMixtureModels2006}.

At iteration $t$ of AMIS, we seek to minimize $\E[\trg]{-\log q^\text{pl}_{t+1}(\param)}$ with respect to the parameters of $q^\text{pl}_{t+1}$, however, the preceding samples $\params^{(1)},\ldots,\params^{(t)}$ are drawn from $q_1,\ldots,q_t$ instead of the posterior $\trg$. As such, we use a weighted EM algorithm~\cite{cornuetAdaptiveMultiple2012} to minimize
\begin{equation}\label{eq:wtd_EM_obj}
    - \frac{1}{N_1+\cdots+N_t}\sum_{s=1}^t \sum_{n=1}^{N_s} \frac{\unnormtrg(\param^{(s)}_n)}{q_{1{:}t}(\param^{(s)}_n)} \log q^\text{pl}_{t+1}(\param^{(s)}_n),
\end{equation}
where
\begin{equation}\label{eq:q1t_defn}
    q_{1{:}t}(\param) = \frac{1}{N_1+\cdots+N_t} \sum_{s=1}^t N_s q_s(\param),
\end{equation}
is the distribution of the collection of samples $\params^{(1)},\ldots,\params^{(t)}$ under a deterministic multiple mixture interpretation~\cite{veachOptimallyCombining1995,owenSafeEffective2000}. The term $\unnormtrg/q_{1{:}t}$ in \eqref{eq:wtd_EM_obj} accounts for the fact that the samples are drawn from $q_{1{:}t}$ instead of the posterior $\trg$. Note that since $\unnormtrg = Z\trg$, the expression in \eqref{eq:wtd_EM_obj} is an estimate of $Z\,\E[\trg]{-\log q^\text{pl}_{t+1}(\param)}$. The weighted EM algorithm is used to find the mixture parameters of
\begin{equation}\label{eq:gmix_pl}
    q^\text{pl}_{t+1}(\param) = \sum_{k=1}^{K^\text{pl}} \gmixw_k^{(t+1)} \phi\!\left(\param; \mu_k^{(t+1)}, \bm\Sigma_k^{(t+1)}\right),
\end{equation}
by repeating, for each $k=1,\ldots,K^\text{pl}$, the updates
\begin{align*}
    \resp^{(s)}_{n,k} &\leftarrow \frac{\unnormtrg(\param^{(s)}_n)}{q_{1{:}t}(\param^{(s)}_n)} \frac{\gmixw^{(t+1)}_k \phi\!\left(\param^{(s)}_n; \mu^{(t+1)}_k, \bm\Sigma^{(t+1)}_k\right)}{\sum_{j=1}^{K^\text{pl}} \gmixw^{(t+1)}_j \phi\!\left(\param^{(s)}_n; \mu^{(t+1)}_j, \bm\Sigma^{(t+1)}_j\right)} \qquad \text{for } s=1,\ldots,t \text{ and } n=1,\ldots,N_s, \\
    \gmixw^{(t+1)}_k &\leftarrow \frac{\sum_{s=1}^t \sum_{n=1}^{N_s} \resp^{(s)}_{n,k}}{\sum_{s=1}^t \sum_{n=1}^{N_s} \sum_{j=1}^{K^\text{pl}} \resp^{(s)}_{n,j}} , \\ 
    \mu^{(t+1)}_k &\leftarrow \frac{\sum_{s=1}^t \sum_{n=1}^{N_s} \resp^{(s)}_{n,k} \param^{(s)}_n}{\sum_{s=1}^t \sum_{n=1}^{N_s} \resp^{(s)}_{n,k}}, \\ 
    \bm\Sigma^{(t+1)}_k &\leftarrow \frac{\sum_{s=1}^t \sum_{n=1}^{N_s} \resp^{(s)}_{n,k} \left(\param^{(s)}_n-\mu^{(t+1)}_k\right)\left(\param^{(s)}_n-\mu^{(t+1)}_k\right)^\trp}{\sum_{s=1}^t \sum_{n=1}^{N_s} \resp^{(s)}_{n,k}}.
\end{align*}
 The only substantive difference between the standard EM updates and the weighted EM updates is the inclusion of the $\unnormtrg/q_{1{:}t}$ term in the responsibilities. In other words, the responsibilities of each sample are adjusted to account for the fact that the samples $\params^{(1)},\ldots,\params^{(t)}$ used in the EM updates are drawn from $q_{1{:}t}$ instead of the posterior $\trg$.

We terminate the weighted EM algorithm after 100 update iterations, or when the relative change in the objective (shown in \eqref{eq:wtd_EM_obj}) is less than $10^{-8}$. The number of mixture components in the placeholder proposal $q^\text{pl}_{t+1}$ is arbitrarily set to $K^\text{pl}=50$, which we consider to be large enough to approximate the posterior distribution. Larger values of $K^\text{pl}$ allow for more flexible proposal distributions at the cost of increased runtime. 

\subsection{Using Pathfinder to initialize robust AMIS}\label{sec:pathfinder}

In robust AMIS, we use a Gaussian mixture as the initial proposal distribution $q_1$ instead of the Laplace IS proposal distribution; the construction of $q_1$ is described in Algorithm~\ref{alg:init_rAMIS}. The components of this Gaussian mixture are selected from local Gaussian approximations of the posterior $\trg$ returned by the Pathfinder algorithm~\cite{zhangPathfinderParallel2022}. The Pathfinder algorithm minimizes $-\log \unnormtrg(\param)$ with respect to $\param$ and constructs approximate Hessians of $-\log \unnormtrg(\param)$ along the optimization path, which can be interpreted as local Gaussian approximations of $\trg$. The authors of Pathfinder recommend performing $I=20$ runs of Pathfinder~\cite{zhangPathfinderParallel2022}, we instead perform $I=50$ runs of Pathfinder as it results in reduced variance in robust AMIS estimates. Let $K^\text{PF}$ be the number of local Gaussian approximations obtained over $I$ runs of Pathfinder, which we denote as $\{\ndist{\mu^\text{PF}_k}{\bm\Sigma^\text{PF}_k}\}_{k=1}^{K^\text{PF}}$. The initial proposal is a uniform mixture of only a subset of these Gaussians, namely
\begin{equation}\label{eq:init_prop}
    q_1(\param) = \sum_{k\in \mathcal{K}^\text{init}} \frac{1}{\lvert \mathcal{K}^\text{init} \rvert} \phi(\param; \mu^\text{PF}_k, \bm\Sigma^\text{PF}_k),
\end{equation}
where $\mathcal{K}^\text{init}$ is a subset of $\{1,\ldots,K^\text{PF}\}$. To select $\mathcal{K}^\text{init}$, we first remove Gaussians that are unlikely to capture the typical set of the posterior $\trg$ well, then greedily select Gaussians that are sufficiently dissimilar from one another, as measured by their pairwise Hellinger distances (defined in \eqref{eq:sq_hellinger}).

\begin{algorithm}[t!]
\caption{Proposal initialization in robust AMIS}\label{alg:init_rAMIS}
\begin{algorithmic}[1]
\Require Unnormalized posterior $\unnormtrg$, prior means $\mu^\text{prior}$, prior standard deviations $\sigma^\text{prior}$, $K^\text{PF}$ local Gaussian approximations $\{\ndist{\mu^\text{PF}_k}{\bm\Sigma^\text{PF}_k}\}_{k=1}^{K^\text{PF}}$ found by Pathfinder, MAP estimate $\MAP$ found by Pathfinder, squared Hellinger distance threshold $\delta$
\State $\mathcal{K}^* \gets \{\}$
\For{$k = 1, \ldots, K^\text{PF}$}
    \If{$\log \unnormtrg(\mu^\text{PF}_k) > \log \unnormtrg(\MAP) - 2d$ and the criteria in \eqref{eq:crit} are met}
        \State $\mathcal{K}^* \gets \mathcal{K}^* \cup \{k\}$
    \EndIf
\EndFor
\State $\mathcal{K}^\text{init} \gets \{\}$
\For{$i = 1, \ldots, \lvert \mathcal{K}^*\rvert$}
    \State $k^* \gets $ index in $\mathcal{K}^*$ with $i$-th highest $\unnormtrg(\mu^\text{PF}_{k^*})$
    \For{$k \in \mathcal{K}^\text{init}$}
        \If{$H^2(\ndist{\mu^\text{PF}_k}{\bm\Sigma^\text{PF}_k},\ndist{\mu^\text{PF}_{k^*}}{\bm\Sigma^\text{PF}_{k^*}}) \le \delta$} \Comment{$H^2$ defined in \eqref{eq:sq_hellinger}}
            \State \textbf{go to} line 9 with next $i$
        \EndIf        
    \EndFor
    \State $\mathcal{K}^\text{init} \gets \mathcal{K}^\text{init} \cup \{k\}$
\EndFor
\State \Return $q_1(\param) = \sum_{k\in \mathcal{K}^\text{init}} \frac{1}{\lvert \mathcal{K}^\text{init} \rvert} \phi(\param; \mu^\text{PF}_k, \bm\Sigma^\text{PF}_k)$
\end{algorithmic}
\end{algorithm}

For each $k=1,\ldots,K^\text{PF}$, we remove the Gaussian $\ndist{\mu^\text{PF}_k}{\bm\Sigma^\text{PF}_k}$ if it does not meet all of the following criteria:
\begin{itemize}
    \item the unnormalized log-posterior at the mean is not too low relative to the maximum found by Pathfinder, i.e. $\log \unnormtrg(\mu^\text{PF}_k) > \log \unnormtrg(\MAP) - 2d$, where $\MAP$ is the maximizer of $\log \unnormtrg$ found by Pathfinder and $d$ is the dimensionality of $\param$,
    \item for each entry of $\param \sim \ndist{\mu^\text{PF}_k}{\bm\Sigma^\text{PF}_k}$, its variance is smaller than the corresponding prior variance,
    \item for each entry of $\param \sim \ndist{\mu^\text{PF}_k}{\bm\Sigma^\text{PF}_k}$, its mean differs from the corresponding prior mean by less than four times the corresponding prior standard deviation.
\end{itemize}
The last two criteria can be expressed as
\begin{equation}\label{eq:crit}
     [\bm\Sigma^\text{PF}_k ] _{ii} < [\sigma^\text{prior}]_i^2 \quad \text{and} \quad \lvert [\mu_k^\text{PF}]_i - [\mu^\text{prior}]_i \rvert < 4[\sigma^\text{prior}]_i \quad \text{for } i =1,\ldots,d,
\end{equation}
where $[\cdot]_i$ denotes the $i$-th entry of a vector, $[\cdot]_{ii}$ denotes the $i$-th entry of the main diagonal of a matrix, $\mu^\text{prior}$ is the vector of prior means of $\param$, and $\sigma^\text{prior}$ is the vector of prior standard deviations of $\param$. The first criterion excludes Gaussians whose centers have relatively low posterior density; the second criterion excludes Gaussians that are too broad relative to the prior distribution; the third criterion excludes Gaussians whose centers are far from the typical set of the prior distribution. 

Let $\mathcal{K}^*$ denote the indices of the remaining Gaussians. The indices corresponding to the Gaussians used in the initial proposal (see \eqref{eq:init_prop}) is a subset of $\mathcal{K}^*$, where the selected Gaussians are sufficiently dissimilar according to a squared Hellinger distance threshold. The squared Hellinger distance between two multivariate Gaussians $\mathcal{N}(\mu,\bm\Sigma)$ and $\mathcal{N}(\mu',\bm\Sigma')$ is given by
\begin{align}
    H^2\!\left(\mathcal{N}(\mu,\bm\Sigma),\mathcal{N}(\mu',\bm\Sigma')\right) 
    &= \frac{1}{2} \int \left( \sqrt{\phi(\param;\mu,\bm\Sigma)} - \sqrt{\phi(\param;\mu',\bm\Sigma')} \right)^2 \,\dd{\param} \nonumber\\
    &= 1 - \frac{\det(\bm\Sigma)^{1/4}\det(\bm\Sigma')^{1/4}}{\det\!\left(\frac{\bm\Sigma+\bm\Sigma'}{2}\right)^{1/2}} \exp\!\left(-\frac{1}{8}(\mu-\mu')^\top\!\left(\frac{\bm\Sigma+\bm\Sigma'}{2}\right)^{\!-1}(\mu-\mu')\right), \label{eq:sq_hellinger}
\end{align}
and is bounded between zero and unity; see~\cite{pardoDivergenceMeasures2006} for a derivation. We build up $\mathcal{K}^\text{init}$ incrementally by processing indices $k^*\in \mathcal{K}^*$ in decreasing order of $\unnormtrg(\mu^\text{PF}_{k^*})$. The index $k^*$ is added into $\mathcal{K}^\text{init}$ if the squared Hellinger distances between $\ndist{\mu^\text{PF}_{k^*}}{\bm\Sigma^\text{PF}_{k^*}}$ and each Gaussian selected so far, i.e. $\{\ndist{\mu^\text{PF}_{k}}{\bm\Sigma^\text{PF}_{k}}\}_{k\in\mathcal{K}^\text{init}}$, are all greater than some threshold $\delta$. In this work, we use a threshold of $\delta=0.1$. Larger values of $\delta$ lead to less Gaussian components used in the initial proposal, which may compromise its coverage of the posterior; smaller values of $\delta$ increases computational burden as more components are included in the initial proposal.

\subsection{Variance reduction in robust AMIS}\label{sec:var_reduction}

The Gaussian mixture proposals $\{q^\text{pl}_{t}\}_{t=2}^{T}$ constructed in standard AMIS are designed to minimize $\E[\trg]{-\log q^\text{pl}_{t}(\param)}$. In robust AMIS, we reduce the variance of the evidence estimate by subsequently minimizing an alternative objective, namely $\E[\trg]{\unnormtrg(\param)/q^\text{ov}_t(\param)}$, with respect to the Gaussian mixture weights, where $q^\text{ov}_t$ is an overall proposal which aims to represent all $N_1+\cdots+N_T$ samples, which we define later in~\eqref{eq:rAMIS_proposal}. To understand why minimizing this alternative objective is related to variance reduction, consider the standard importance sampling estimator
\begin{equation*}
    \hat{Z} = \frac{1}{N} \sum_{n=1}^N \frac{\unnormtrg(\param_n)}{q(\param_n)},
\end{equation*}
where $q$ is a generic proposal from which $N$ samples $\param_1,\ldots,\param_N$ are drawn independently. Since
\begin{equation*}
    \E[q]{\frac{\unnormtrg(\param)}{q(\param)}} = \int q(\param) \frac{\unnormtrg(\param)}{q(\param)} \, \dd{\param} = \int \unnormtrg(\param) \, \dd{\param} = Z,
\end{equation*}
it follows that $\E[q]{\hat{Z}} = Z$, i.e. the estimator $\hat{Z}$ is unbiased. The variance of $\hat{Z}$ can be written as
\begin{align}
    \mathrm{Var}(\hat{Z}) 
    &= \frac{1}{N} \Var[q]{\frac{\unnormtrg(\param)}{q(\param)}} \nonumber\\
    &= \frac{1}{N} \left( \E[q]{\frac{\unnormtrg(\param)^2}{q(\param)^2}} - \E[q]{\frac{\unnormtrg(\param)}{q(\param)}}^2 \right) \nonumber\\
    &= \frac{1}{N} \left( \int q(\param) \frac{\unnormtrg(\param)^2}{q(\param)^2} \, \dd{\param} - Z^2 \right) \nonumber\\
    &= \frac{1}{N} \left( \int Z\, \trg(\param) \frac{\unnormtrg(\param)}{q(\param)} \, \dd{\param} - Z^2 \right) \nonumber\\
    &= \frac{1}{N}\left(Z \, \E[\trg]{\frac{\unnormtrg(\param)}{q(\param)}}  - Z^2\right). \label{eq:IS_var}
\end{align}
This shows that for standard importance sampling, minimizing the variance of $\hat{Z}$ is equivalent to minimizing $\E[\trg]{\unnormtrg(\param)/q(\param)}$. The derivation in \eqref{eq:IS_var} requires samples $\param_1,\ldots,\param_N$ to be independent, a condition that is generally not true for AMIS due to proposal adaptation. Nevertheless, for each $t=1,\ldots,T-1$, robust AMIS adjusts the Gaussian mixture weights of $q^\text{pl}_{t+1}$ obtained with the weighted EM algorithm (Section~\ref{sec:wem}) by minimizing an estimate of $\E[\trg]{\unnormtrg(\param)/q^\text{ov}_{t+1}(\param)}$ based on the existing samples $\param^{(1)},\ldots,\param^{(t)}$, where $q^\text{ov}_{t+1}$ is an overall proposal distribution. The overall proposal $q^\text{ov}_{t+1}$ is constructed under the deterministic multiple mixture interpretation~\cite{veachOptimallyCombining1995,owenSafeEffective2000}, assuming that the collection of all $N_1+\cdots+N_T$ samples consists of $N_s$ samples drawn from $q_s$ for each $s=1,\ldots,t$ and $N_{t+1}+\cdots+N_T$ samples drawn from $q^\text{pl}_{t+1}$. This results in the overall proposal
\begin{equation}\label{eq:rAMIS_proposal}
    q^\text{ov}_{t+1}(\param) = \sum_{s=1}^{t} \frac{N_s}{N_1 + \cdots + N_T} q_s(\param) + \frac{N_{t+1} + \cdots + N_T}{N_1 + \cdots + N_T} q^\text{pl}_{t+1}(\param).
\end{equation}

We seek to minimize, with respect to the Gaussian mixture weights $\{\gmixw_k^{(t+1)}\}_{k=1}^{K^\text{pl}}$, the objective
\begin{equation}\label{eq:rAMIS_obj}
\frac{1}{N_1+\cdots+N_t}\sum_{s=1}^t \sum_{n=1}^{N_s} \frac{\unnormtrg(\param^{(s)}_n)}{q_{1{:}t}(\param^{(s)}_n)} \frac{\unnormtrg(\param^{(s)}_n)}{q^\text{ov}_{t+1}(\param^{(s)}_n)}, 
\end{equation}
where $q_{1{:}t}$ is defined in \eqref{eq:q1t_defn}. Similar to the weighted EM algorithm, the term $\unnormtrg/q_{1{:}t}$ in \eqref{eq:rAMIS_obj} accounts for the fact that the samples $\params^{(1)},\ldots,\params^{(t)}$ are drawn from $q_1,\ldots,q_{t}$ instead of $\trg$. With the values of $\{(\mu_k^{(t+1)}, \bm\Sigma_k^{(t+1)})\}_{k=1}^{K^\text{pl}}$ fixed from running the weighted EM algorithm, the overall proposal $q^\text{ov}_{t+1}(\param)$ in \eqref{eq:rAMIS_proposal} is an affine expression with respect to $\{\gmixw_k^{(t+1)}\}_{k=1}^{K^\text{pl}}$ (see \eqref{eq:gmix_pl}). Since the function $x\mapsto 1/x$ is convex in $x$, it follows that the objective in \eqref{eq:rAMIS_obj} to be minimized is convex in $\{\gmixw_k^{(t+1)}\}_{k=1}^{K^\text{pl}}$. This minimization problem is defined over a simplex, as $\{\gmixw_k^{(t+1)}\}_{k=1}^{K^\text{pl}}$ are nonnegative reals that sum to unity. We perform minimization using the Cauchy-Simplex algorithm~\cite{chokOptimizationProbability2025}, which solves convex minimization problems over a simplex.

\section{Bridge sampling}\label{sec:bridge_supp}

Bridge sampling estimates model evidence by utilizing samples drawn from a proposal distribution~$q$ and samples drawn from the posterior distribution~$\trg$. Compared to performing importance sampling with samples drawn from $q$ alone, the bridge sampling estimate is less sensitive to the mismatch between $q$ and $\trg$ due to the use of a bridge function (\eqref{eq:optimal_BS} and \eqref{eq:iterative_BS}) that mediates between $q$ and $\trg$. 

\subsection{Theory}

Let $q(\param)$ be a generic proposal density. For any function $\alpha(\param)$ defined on the intersection of the supports of $\pi$ and $q$, the bridge sampling identity states that
\begin{equation}\label{eq:BS_identity}
    \frac{\E[q]{\unnormtrg(\param)\alpha(\param)}}{\E[\trg]{q(\param)\alpha(\param)}} = \frac{\int q(\param)\unnormtrg(\param)\alpha(\param) \, \dd{\param}}{\int \trg(\param)q(\param)\alpha(\param) \, \dd{\param}} = \frac{Z \int q(\param)\trg(\param)\alpha(\param) \, \dd{\param}}{\int \trg(\param)q(\param)\alpha(\param) \, \dd{\param}} = Z.
\end{equation}
The function $\alpha$ is referred to as the \emph{bridge function}. Given $N_q$ samples $\param^q_1,\ldots,\param^q_{N_q}$ drawn from the proposal~$q$ and $N_\trg$ samples $\param^\trg_1,\ldots,\param^\trg_{N_\trg}$ drawn from the posterior~$\trg$, the bridge sampling estimate of $Z$ is
\begin{equation}\label{eq:BS_estimate}
    \hat{Z} = \frac{N_q^{-1} \sum\limits_{n=1}^{N_q} \unnormtrg(\param^q_n)\alpha(\param^q_n)}{N_\trg^{-1}\sum\limits_{n=1}^{N_\trg} q(\param^\trg_n)\alpha(\param^\trg_n)}.
\end{equation}
Assuming that the posterior samples are independently drawn (and likewise for the proposal samples), \textcite{mengSimulatingRatios1996a} show that the bridge function
\begin{equation}\label{eq:optimal_BS}
    \alpha_\text{opt}(\param) \propto \frac{1}{N_q q(\param) + N_\trg \trg(\param)},
\end{equation}
minimizes a first-order approximation of the relative error of $\hat{Z}$. Note that the bridge function only needs to be defined up to a multiplicative constant; see \eqref{eq:BS_estimate}. However, evaluating the bridge function $\alpha_\text{opt}$ requires evaluating the posterior density $\trg$, i.e. the value of $Z$ is needed. In response, \textcite{mengSimulatingRatios1996a} propose an iterative scheme: given some current estimate $\hat{Z}$, the bridge function is updated to
\begin{equation}\label{eq:iterative_BS}
    \alpha_\text{opt}(\param; \hat{Z}) \propto \frac{1}{N_q q(\param) + N_\trg \unnormtrg(\param)/\hat{Z}},
\end{equation}
while the estimate $\hat{Z}$ is updated according to \eqref{eq:BS_estimate} with $\alpha(\param) = \alpha_\text{opt}(\param; \hat{Z})$.

\subsection{Implementation}

We use the No-U-Turn Sampler (NUTS)~\cite{hoffmanNoUturnSampler2014} to obtain posterior samples for bridge sampling and follow the iterative scheme of \textcite{mengSimulatingRatios1996a}. We use $10^4$ posterior samples to construct the proposal distribution $q$; the remaining posterior samples are used to estimate $\E[\trg]{q(\param)\alpha(\param)}$, i.e. compute the denominator of \eqref{eq:BS_estimate}. The total number of samples used for each example is reported in Section~5 of the main text. Since NUTS produces autocorrelated samples, the independence assumption used to derive the optimal bridge function $\alpha_\text{opt}$ in \eqref{eq:optimal_BS} is violated. Instead, we replace $N_\trg$ in \eqref{eq:iterative_BS} with the rank-normalized effective sample size~\cite{vehtariRanknormalizationFolding2021} of the $N_\trg$ samples used to estimate $\E[\trg]{q(\param)\alpha_\text{opt}(\param;\hat{Z})}$, as suggested in \cite{fruhwirth-schnatterEstimatingMarginal2004}.

We specify the proposal $q$ to be a Gaussian mixture distribution of $K=50$ components. The parameters of the Gaussian mixture are found using the standard EM algorithm; the weighted EM algorithm is not needed as we fit the Gaussian mixture using samples drawn from the posterior distribution~$\trg$.

\section{Datasets and statistical models}\label{sec:ode_supp}

In this section, we present details pertaining the datasets, likelihood, and parameter priors used in the examples of the main text.

\subsection{Coral re-growth example}\label{sec:coral_supp}

We consider the following models for the growth of a single coral population $C$ over time $t$:
\begin{align}
    \frac{\dd{C}}{\dd{t}} &= rC\left(1-\frac{C}{K}\right), \tag{Logistic} \\
    \frac{\dd{C}}{\dd{t}} &= rC\log\!\left(\frac{K}{C}\right), \tag{Gompertz} \\
    \frac{\dd{C}}{\dd{t}} &= rC\left[1-\left(\frac{C}{K}\right)^\beta\right]. \tag{Richards'}    
\end{align}

\begin{figure}[ht]
    \centering
    \begin{minipage}{\linewidth}
        \centering
        \includegraphics[width = 0.55\textwidth]{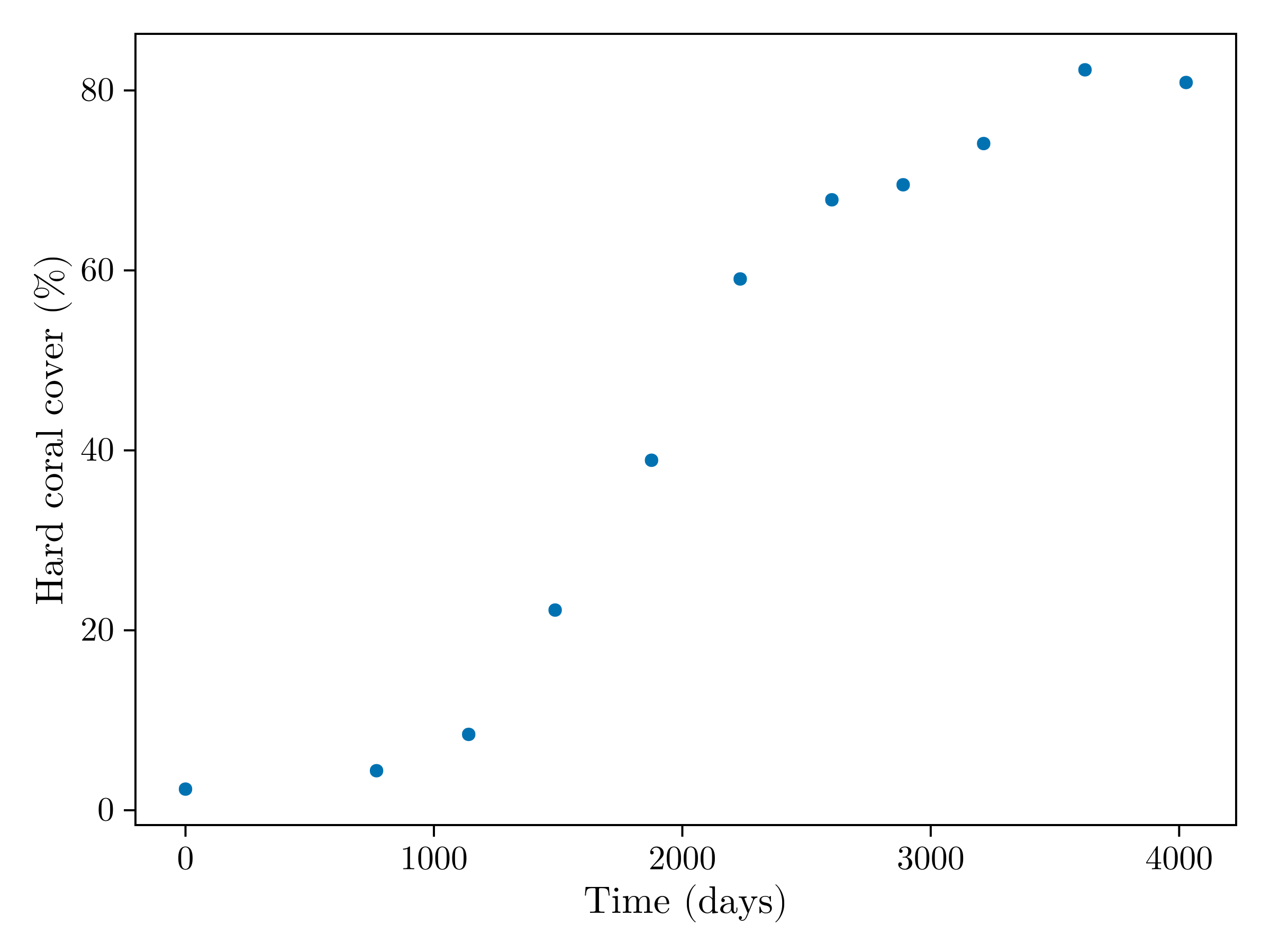}
        \caption{Data for the coral re-growth example showing the percentage area covered by hard coral over time as reported in \cite{simpsonParameterIdentifiability2022}.}
        \label{fig:coral_data}
    \end{minipage}

    \vspace{1em}
    
    \begin{minipage}{\linewidth}
        \centering
        \includegraphics[width = 0.65\textwidth]{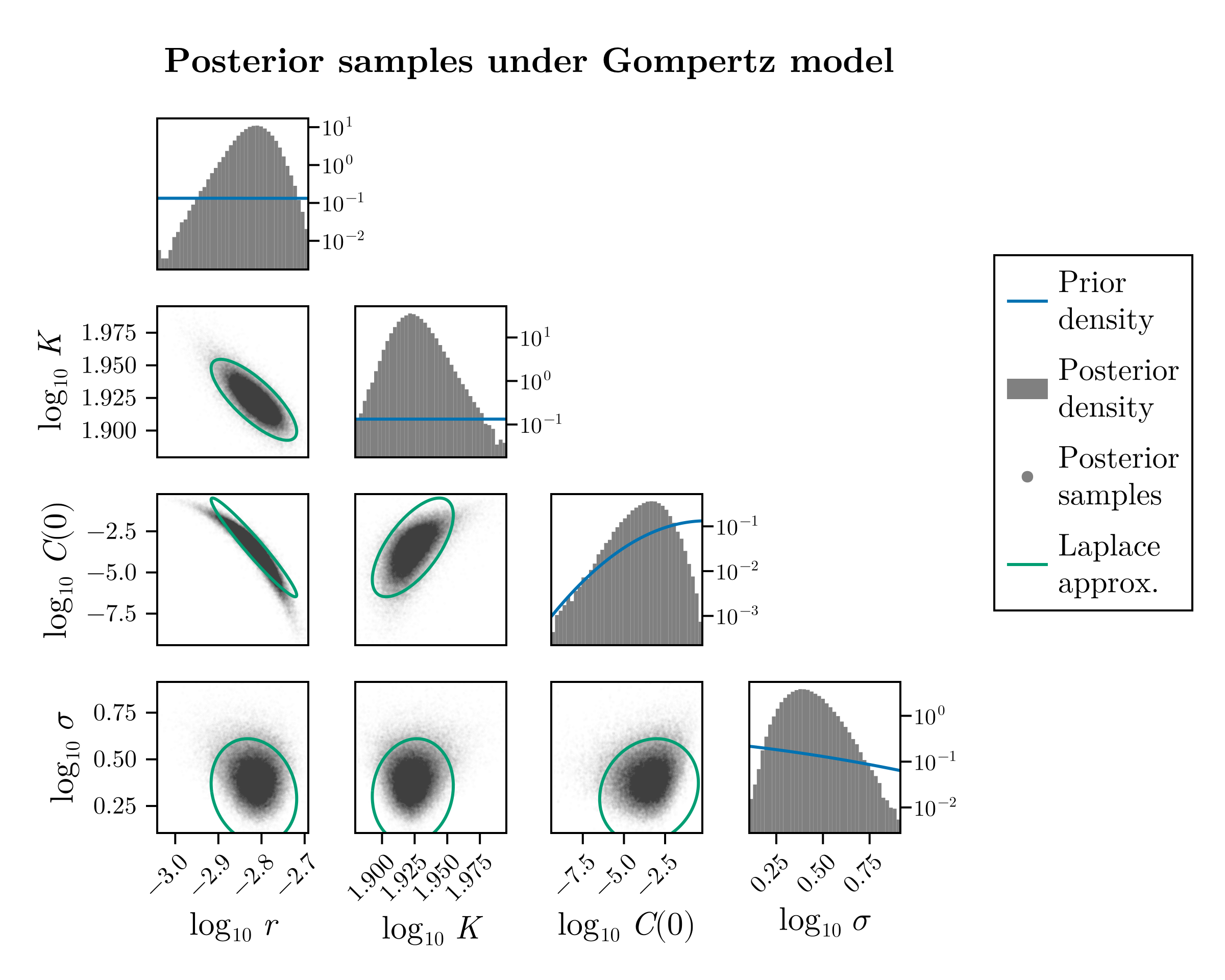}
        \caption{Posterior samples obtained from one NUTS run consisting of five independent chains, under the Gompertz model. Plots along the main diagonal show univariate marginals of the posterior (histogram) along with the priors (blue lines), where density is shown on a log scale. Plots off the main diagonal show bivariate marginals of the posterior (gray points) along with the 95\% probability region of the Gaussian approximation centered at the MAP estimate (green ellipses).}
        \label{fig:gompertz}
    \end{minipage}
\end{figure}

We use the dataset reported in \cite{simpsonParameterIdentifiability2022}, which consists of $S=11$ observations $y_1,\cdots,y_S$ of hard coral cover~(\%) at times $t_1,\cdots,t_S$~(days); see plotted data in Figure~\ref{fig:coral_data}. The data describes the re-growth of hard coral cover on a reef near Lady Musgrave Island, Australia, after some external disturbance (such as a tropical cyclone). Following \cite{simpsonParameterIdentifiability2022}, we specify a likelihood assuming that observation noise is identically and independently distributed, additive, and follows a normal distribution:
\begin{equation*}
    y_i \vert \param \sim \mathcal{N}(C(t_i), \sigma^2) \quad \text{for } i=1,\ldots,S. 
\end{equation*}
For the logistic and Gompertz models, the parameter vector is $\param = (\log_{10} r, \log_{10} K, \log_{10} C(0), \log_{10} \sigma)$. The Richards' model additionally includes $\log_{10} \beta$ as a parameter. The parameter priors are as follows:
\begin{align*}
    \log_{10} r &\sim \mathcal{N}(-3, 3), \\
    \log_{10} K &\sim \mathcal{N}(2, 3), \\
    \log_{10} C(0) &\sim \mathcal{N}(0, 3), \\
    \log_{10} \beta &\sim \mathcal{N}(0, 3), \\
    \log_{10} \sigma &\sim \mathcal{N}(0, 1). 
\end{align*}
All parameter priors are shared across models, except that the prior for $\log_{10} \beta$ is used only by the Richards' model. All priors are weakly informative as the prior standard deviation corresponds to three orders of magnitude, except for the noise standard deviation $\sigma$, which is given a narrower prior. The prior for $\sigma$ is chosen to discourage large values of $\sigma$ that would lead to underfitting; see the prior probability interval of $\sigma$ in the final item of the list below. The prior means are informed by domain knowledge about coral re-growth:
\begin{itemize}
    \item the prior mean of the log-transformed growth rate $\log_{10} r$ suggests a timescale of 1000 days (c.f. observation duration of 4028 days, Figure~\ref{fig:coral_data});
    \item the prior mean of the log-transformed carrying capacity $\log_{10} K$ matches the maximum possible coral cover, namely $K=100\%$;
    \item the prior mean of the log-transformed initial population $\log_{10} C(0)$ suggests an initial coral cover of 1\%;
    \item the prior mean of $\log_{10} \beta$ corresponds to $\beta=1$, which reduces the Richards' model to the logistic model;
    \item the prior mean of the log-transformed noise standard deviation $\log_{10} \sigma$ corresponds to a 95\% prior probability that $\sigma \in [0.01\%, 100\%]$.
\end{itemize}

\subsection{Insect life-stage example}\label{sec:insect_supp}

We consider the following system of differential equations describing the population dynamics of three life stages of a hypothetical insect, namely egg $(E)$, larva $(L)$, and adult $(A)$:
\begin{align*}
    \frac{\dd{E}}{\dd{t}} &= \rho A - \lambda_{EL} E - \delta_E E - \frac{1}{2}\kappa_E E^2, \\
    \frac{\dd{L}}{\dd{t}} &= \lambda_{EL} E - \lambda_{LA} L - \delta_L L - \frac{1}{2}\kappa_L L^2, \\
    \frac{\dd{A}}{\dd{t}} &= \lambda_{LA} L - \delta_A A - \frac{1}{2}\kappa_A A^2.
\end{align*}
The differential equation parameters consist of the reproduction rate~$\rho$, the inverse durations of the egg and larva stages~$\lambda_{EL},\lambda_{LA}$, the first-order death rates of each stage~$\delta_E,\delta_L,\delta_A$, and the second-order death rates of each stage~$\kappa_E,\kappa_L,\kappa_A$. The factor of $1/2$ arises from applying mass-action kinetics to events where two individuals of the same population compartment interact with each other. Note that this factor would not appear for competition between different population compartments. Here, we used the chemical reaction network modeling software package \texttt{Catalyst.jl}~\cite{lomanCatalystFast2023} to model population dynamics. The \texttt{Catalyst.jl} package automatically includes the aforementioned combinatorial factors for reactions involving multiple individuals from the same population compartment. We consider the reproduction and stage transitions to be core mechanisms. A set~$\modelspace$ of $M=64$ models is generated by setting each possible subset of $\{\delta_E,\delta_L,\delta_A,\kappa_E,\kappa_L,\kappa_A\}$ to be zero. In other words, each model corresponds to a subset of the death mechanisms being present in the population.

We simulate synthetic datasets each consisting of $S=41$ observations $\textbf{y}_1,\ldots,\textbf{y}_S$ of $(E,L,A)$ at times $t_1,\ldots,t_S$ equally spaced across the interval $[0, 10]$ (all quantities given in arbitrary units). For each model, the parameter vector $\param$ consists of the $\log_{10}$-transformed rates of the core mechanisms, the $\log_{10}$-transformed rates of the corresponding subset of death mechanisms, and $\log_{10} \sigma$, where $\sigma$ is a noise parameter. The initial condition is assumed to be known, and is fixed at $(E(0),L(0),A(0))=(0,0,3)$ for all datasets. We assume the noise model to be a mixture of additive and multiplicative noise:
\begin{equation*}
    y_{ij} \vert \param \sim \mathcal{N}(x_j(t_i), (0.01+\sigma x_j(t_i))^2) \quad \text{for } i=1,\ldots,S, \, j=1,2,3,
\end{equation*}
where $\textbf{y}_i=(y_{i1},y_{i2},y_{i3})$ for each $i=1,\ldots,S$ and $(x_1,x_2,x_3)=(E,L,A)$. Note that the variance of additive noise is fixed, and the multiplicative noise level $\sigma$ is shared across $j=1,2,3$. The noise model is used for simulation and inference; we fix the multiplicative noise level to be $\sigma=0.05$ for all simulations. The noise level is given a prior of $\log_{10} \sigma \sim \mathcal{N}(-1, 1)$, which assumes 1\%--100\% as a plausible range for the multiplicative noise. The priors of all $\log_{10}$-transformed rates are set to $\mathcal{N}(0, 2)$, where the prior mean is informed by the inverse duration of the observation duration, namely $[0,10]$.

\begin{figure}[ht]
\centering
\includegraphics[width = 0.99\textwidth]{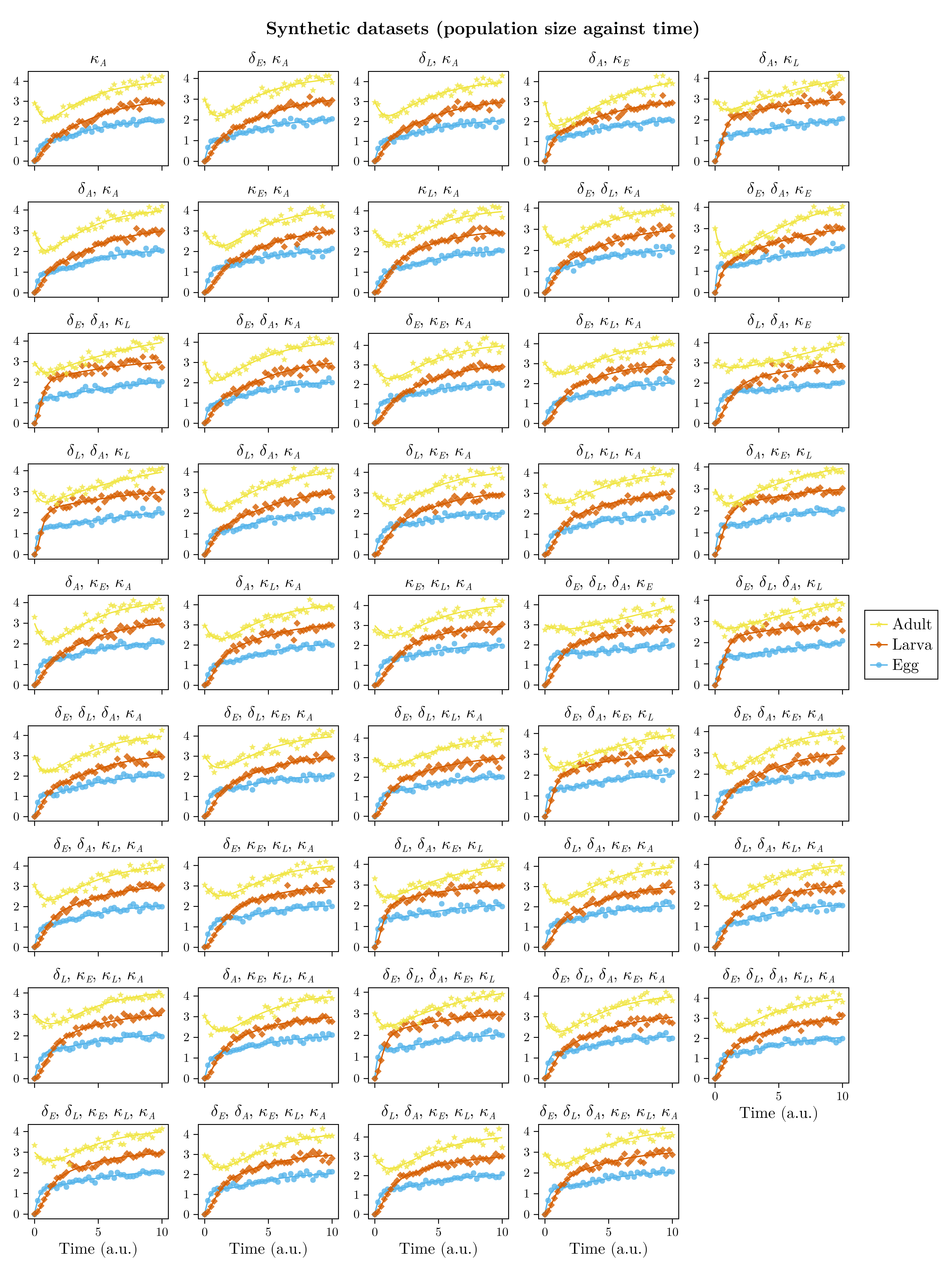}
\caption{Synthetic datasets used for the insect life-stage example. Population size and time are given in arbitrary units. The title above each plot lists the death rate parameters included in the corresponding data-generating model.}
\label{fig:insect_datasets}
\end{figure}

We design the data simulation process to generate datasets of similar trajectories that tend to a positive equilibrium for all life stages (Figure~\ref{fig:insect_datasets}). Each of the 44 simulated datasets correspong to a model in $\modelspace$ that supports a positive equilibrium for all entries of $\mathbf{x}=(x_1,x_2,x_3)$. For a model to support a positive equilibrium, it must include at least one death mechanism for the adult life stage to prevent $A(t)$ from growing unboundedly, and it must include at least one second-order death rate since the only equilibrium of a linear system of homogeneous differential equations is zero. These criteria exclude 20 of the 64 models in $\modelspace$. Although we do not have analytical results that the remaining 44 models have a stable positive equilibrium, we numerically verify that the parameters used to generate each dataset lead to a positive equilibrium where all eigenvalues of the Jacobian of the differential equation system are negative.

The data-generating rate parameters are chosen by minimizing an objective function that encourages the population sizes at the final time~$t_S$ to be close to $\mathbf{x}^\text{trg}=(2,3,4)$, the derivatives $\dd{\mathbf{x}}/\dd{t} = (\dd{x_1}/\dd{t},\dd{x_2}/\dd{t},\dd{x_3}/\dd{t})$ to be of small magnitude at the final time $t_S$, and rate parameters to have an order of magnitude close to $10^0$. Specifically, the objective function is
\begin{equation*}
     \log(1 + (10\lVert \mathbf{x}(t_S) - \mathbf{x}^\text{trg}\rVert)^{10}) + \log\!\left( 1 + \left( 10\bigg\lVert\left.\frac{\dd{\mathbf{x}}}{\dd{t}}\right|_{t = t_S}\bigg\rVert \right)^{10} \right) - \sum_{r\in\text{rate parameters}} \log p_{\Gamma(2,2)}(r),
\end{equation*}
where $\mathbf{x}(t_S)$ and $\left.\dd{\mathbf{x}}/\dd{t}\right|_{t = t_S}$ are functions of the rate parameters, $p_{\Gamma(2,2)}$ is the density function of a gamma distribution with shape and rate parameters equal to 2, and $\lVert\cdot\rVert$ denotes the Euclidean norm of a vector.. Note that $u\mapsto\log(1+(10u)^{10})$ is a function that is relatively flat over $u\in[0,0.1]$, and is of order $\bigO{\log u}$ as $u\rightarrow\infty$. 

\clearpage

\section{Supplementary figures}

\begin{figure}[ht]
    \centering
    \includegraphics[width = 0.99\textwidth]{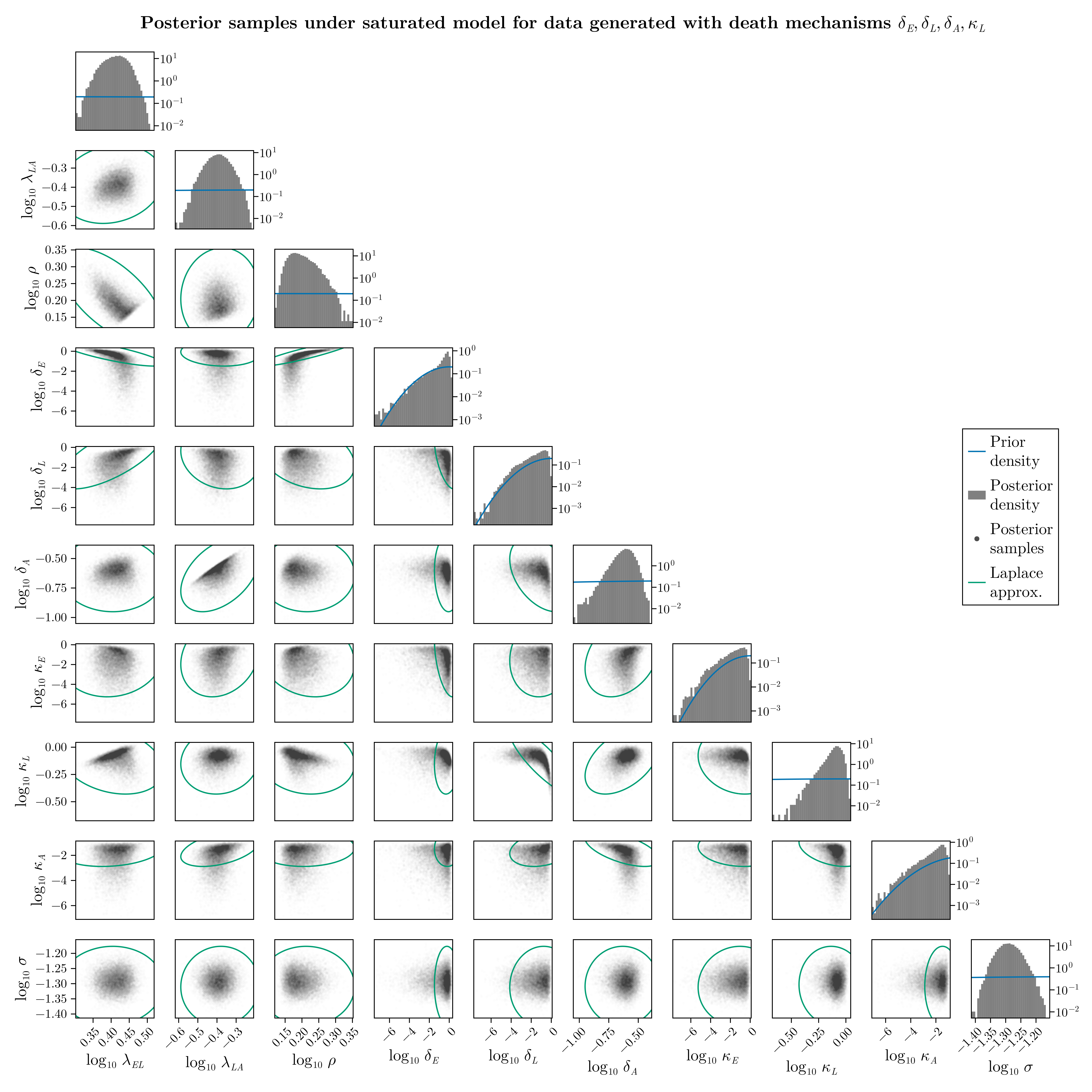}
    \caption{An example of posterior samples obtained with NUTS, under the insect life-stage model with all death mechanisms included. This example uses the same dataset as Figure~4 of the main text, which is generated from the model with death mechanisms corresponding to $\delta_E, \delta_L, \delta_A, \kappa_L$ included. Plots along the main diagonal show univariate marginals of the posterior (histogram) along with the priors (blue lines), where density is shown on a log scale. Plots off the main diagonal show bivariate marginals of the posterior (gray points) along with the 95\% probability region of the Gaussian approximation centered at the MAP estimate (green ellipses).}
    \label{fig:insect_example_post}
\end{figure}


\clearpage

\section{Supplementary tables}

\begin{table}[h]
\centering\small
\begin{tabular}{llcc}
\toprule
& Quantity & \makecell{Notation\\(if present)} & Value \\ 
\midrule
\multicolumn{4}{l}{Laplace IS} \\
& Number of samples drawn from proposal & $N$ & $10^6$ \\ 
\midrule
\multicolumn{4}{l}{Standard AMIS and robust AMIS} \\
& Number of AMIS iterations & $T$ & 16 \\
& Number of samples drawn from initial proposal $q_1$ & $N_1$ & $10^4$ \\
& Number of samples drawn from proposal $q_{t+1}$ for $t<T$ & $N_{t+1}$ & $\lfloor 10^{4+2t/15} \rfloor - \lfloor 10^{4+2(t-1)/15} \rfloor$ \\
& Number of mixture components before truncation & $K^\text{pl}$ & 50 \\
\midrule
\multicolumn{4}{l}{Pathfinder} \\
& Number of Pathfinder runs & $I$ & 50 \\
\midrule
\multicolumn{4}{l}{NUTS} \\
& Number of chains & & $5$ \\ 
& Number of burn-in iterations per chain & & $10^3$ \\ 
& Number of retained samples per chain & & $2\times10^4$ (coral), $3\times10^3$ (insect) \\ 
\midrule
\multicolumn{4}{l}{Bridge sampling} \\
& Number of components in Gaussian mixture proposal & $K$ & 50 \\
& Number of posterior samples used to construct proposal & & $10^4$ \\
& Number of posterior samples used to estimate $\E[\trg]{\cdot}$ & $N_\trg$ & $9\times10^4$ (coral), $5\times10^3$ (insect) \\
& Number of samples drawn from proposal & $N_q$ & $10^6$ \\
\bottomrule
\end{tabular}
\tblspace
\caption{Algorithmic settings for all computational methods. Most values are shared across the coral re-growth and insect life-stage examples; values that differ across the coral and insect examples are labeled accordingly.}\label{tbl:alg_settings}
\end{table}

\begin{table}[h]
\centering\small
\begin{tabular}{lccc}
\toprule
Method & Logistic model & Gompertz model & Richards' model \\
\midrule
Laplace IS & \ftmath{2.49\times 10^{-5} \pm 1.37\times 10^{-3}} & \ftmath{5.50\times 10^{-4} \pm 3.27\times 10^{-3}} & \ftmath{-9.72\times 10^{-1} \pm 6.17\times 10^{-2}} \\
Standard AMIS & \ftmath{-1.22\times 10^{-3} \pm 4.46\times 10^{-4}} & \ftmath{-1.46\times 10^{-3} \pm 3.80\times 10^{-4}} & \ftmath{-3.53\times 10^{-1} \pm 7.52\times 10^{-2}} \\
Robust AMIS & \ftmath{-1.00\times 10^{-3} \pm 3.61\times 10^{-4}} & \ftmath{-1.11\times 10^{-3} \pm 4.19\times 10^{-4}} & \ftmath{-3.03\times 10^{-3} \pm 1.20\times 10^{-2}} \\
Bridge sampling & \ftmath{-1.64\times 10^{-7} \pm 5.76\times 10^{-4}} & \ftmath{-3.54\times 10^{-7} \pm 8.45\times 10^{-4}} & \ftmath{-1.19\times 10^{-6} \pm 1.55\times 10^{-3}} \\
\bottomrule
\end{tabular}
\tblspace
\caption{Summary statistics of the log-evidence errors $\log\hat{Z} - \log Z_\text{gold}$ for the coral re-growth example aggregated over 100 runs of the Monte Carlo methods, formatted as mean $\pm$ standard deviation. Note that $Z_\text{gold}$ is the mean of the bridge sampling evidence estimates.}\label{tbl:comparison_coral}
\end{table}

\clearpage
\AtNextBibliography{\small}
\emergencystretch 2em
\printbibliography[heading=bibintoc]

\end{refsection}

\end{document}